%THIS PAPER USES HARVMAC
%%%%%%%%%%%%%%%%%%%%%%%%%%%%%%%%%%%%%%%%%%%%%%%%%%%%%%%%%%%%%%%%%%%%%%%%%%
\input harvmac
\openup 1 \jot
\def\Title#1#2#3{#3\hfill \break \vskip -0.35in
\rightline{#1}\ifx\answ\bigans\nopagenumbers\pageno0\vskip.2in
\else\pageno1\vskip.2in\fi \centerline{\titlefont #2}\vskip .1in}

%%%%%%%%%%%%%%%%%%
%%%%%%%%%%
%defs

\def\bra#1{\langle #1 |}
\def\ket#1{| #1\rangle}
\def\norm#1{\left\Vert \, #1 \, \right\Vert}
\def\inprod#1#2{\langle \, #1 \, , \, #2 \, \rangle}
\def\S{{\cal{S}}}

\def\R{\hbox{\rm I \kern-5pt R}}

\font\ticp=cmcsc10
\def\ajou#1&#2(#3){\ \sl#1\bf#2\rm(19#3)}
%
%refs
%
%{\baselineskip=24pt
\lref\mcelwaine{J.~McElwaine, {\it ``Approximate and 
Exact Consistency of Histories''},  University of Cambridge
preprint DAMTP/95-32, submitted to
Phys. Rev. A., quant-ph/9506034.}

\lref\whitaker{A.~Whitaker, {\it Einstein, Bohr and the quantum
world}, to be published.}

\lref\cfg{J.~Cushing, A.~Fine and S.~Goldstein (eds), 
{\it Bohmian Mechanics and Quantum Theory: An Appraisal}, 
Kluwer Academic Press, to be published.}

\lref\griff{R.B.~Griffiths, \ajou J. Stat. Phys. &36 (84) 219.}

\lref\grifflogic{R.B.~Griffiths, \ajou Found. Phys. &23 (93) 1601.}

\lref\griffincon{Section 6.2 of \griff.}

\lref\grifftwo{Section 4.3 of \griff.}

\lref\duerretal{D. D\"urr, S. Goldstein and N. Zangh\'\i,
\ajou J. Stat. Phys., &67 (92) 843.}

\lref\zurekptp{W.~Zurek, \ajou Prog. Theor. Phys. &89 (93) 281.}

\lref\pazzurek{J.~Paz and W.~Zurek, \ajou Phys. Rev. D &48 (93) 2728.}

\lref\omnes{R.~Omn\`es, \ajou J. Stat. Phys. &53 (88) 893;
\ajou ibid &53 (88) 933; \ajou ibid &53 (88) 957; \ajou
ibid &57 (89) 357.}

\lref\omnesreview{R.~Omn\`es, \ajou Rev.~Mod.~Phys. &64 (92) 339.}

\lref\omnestrue{Page 366 of \omnesreview.}

\lref\omnesmeasure{Page 362 of \omnesreview.}

\lref\omnesclass{Sections D-F of \omnesreview.}

\lref\omnesnew{R.~Omn\`es, \ajou Phys. Lett. A &187 (94) 26.}

\lref\gmhsantafe{M.~Gell-Mann and J.B.~Hartle in {\it Complexity, Entropy,
and the Physics of Information, SFI Studies in the Sciences of
Complexity}, Vol.
VIII, ed. by W.~Zurek,  Addison Wesley, Reading (1990).}

\lref\gmhone{M.~Gell-Mann and J.B.~Hartle  in {\it Proceedings of
the 3rd
International Symposium on the Foundations of Quantum Mechanics in the
Light of
New Technology} ed.~by S.~Kobayashi, H.~Ezawa, Y.~Murayama,  and
S.~Nomura,
Physical Society of Japan, Tokyo (1990).}

\lref\gmhtwo{M.~Gell-Mann and J.B.~Hartle in {\it Proceedings of
the 25th International Conference on High Energy Physics, Singapore, August
2-8, 1990},
ed.~by K.K.~Phua and Y.~Yamaguchi (South East Asia Theoretical Physics
Association
and Physical Society of Japan) distributed by World Scientific, Singapore
(1990).}

\lref\gmhthree{M.~Gell-Mann and J.B.~Hartle in {\it Proceedings of the
NATO Workshop on the Physical Origins of Time Asymmetry, Mazag\'on, Spain,
September 30-October 4, 1991} ed. by J.~Halliwell, J.~P\'erez-Mercader, and
W.~Zurek, Cambridge University Press, Cambridge (1994), gr-qc/9304023.}

\lref\gmhprd{M.~Gell-Mann and J.B.~Hartle, \ajou Phys. Rev. D
&47 (93) 3345.}

\lref\gmhonmeasure{Pages 448-449 of \gmhsantafe.}

\lref\gmhonbranches{Pages 450-451 of \gmhsantafe; page 3352 of \gmhprd.}

\lref\gmhpathint{Pages 432-434 of \gmhsantafe.}

\lref\gmhquasi{Page 445 of \gmhsantafe.}

\lref\gmhigus{Page 454 of \gmhsantafe.}

\lref\gmhaltern{Page 455 of \gmhsantafe.}

\lref\gmhtriv{Page 441 of \gmhsantafe.}

\lref\gmhcomm{M.~Gell-Mann and J.B.~Hartle, {\it ``Equivalent Sets of
Histories and Multiple Quasiclassical Domains''},  Preprint UCSBTH-94-09,
gr-qc/9404013, submitted to gr-qc 8 April 1994.}

\lref\gmhcommnew{M.~Gell-Mann and J.B.~Hartle, {\it ``Equivalent Sets
of
Histories and Multiple Quasiclassical Domains''},  Preprint
UCSBTH-94-09, revised version as of 26 April 1995.}

\lref\gmhoverlap{Page 16 of \gmhcomm; see also M.~Gell-Mann,
{\it The Quark and the Jaguar}, Little, Brown and Co, London
(1994), pages 164-5.}

\lref\hartleone{J.B.~Hartle, in {\it Quantum Cosmology and Baby
Universes}, Proceedings of the 1989 Jerusalem Winter School on
Theoretical Physics,
ed.~by S.~Coleman, J.~Hartle, T.~Piran, and S.~Weinberg,
World Scientific, Singapore (1991).}

\lref\rhofhartle{See for example Section IV.2 of \hartleone.} 

\lref\hartletwo{J.B.~Hartle, \ajou Phys. Rev. &D44 (91) 3173.}

\lref\hartlethree{J.B.~Hartle, {\it ``Spacetime Quantum Mechanics and the
Quantum Mechanics of Spacetime''} in {\it Gravitation and Quantizations:
Proceedings of the 1992 Les Houches
Summer School}, ed. by B.~Julia and J.~Zinn-Justin, North Holland
Publishing Co, Amsterdam, (1994), gr-qc/9304006.}

\lref\hartlepast{Page 77 of \hartleone.}

\lref\abl{Y.~Aharonov, P.~Bergmann and J.~Lebovitz, \ajou
Phys. Rev. &B134 (64) 1410.} 

\lref\hartleequiv{Pages 25-26 of \hartleone.}

\lref\hartlemarvellous{Section V.1.2 of \hartleone.}

\lref\hartlepriv{J.B.~Hartle, private communication.}

\lref\goldsteinpriv{S.~Goldstein, private communication.}

\lref\omnespriv{R.~Omn\`es, private communication.}

\lref\griffithspriv{R.~Griffiths, private communication.}

\lref\isham{C.J.~Isham,
{\it ``Canonical quantum gravity and the problem of time''} in      
{\it ``Integrable Systems, Quantum Groups, and Quantum Field     
Theories''}, L.~A.~Ibort and M.~A.~Rodriguez (eds.),
Kluwer, London (1993); 
C.J.~Isham, \ajou J. Math. Phys. &23 (94) 2157; 
C.J.~Isham and N.~Linden, \ajou J. Math. Phys. &35 (94) 5452.}

\lref\albrecht{A.~Albrecht, \ajou Phys. Rev. &D46 (92) 5504; \ajou
Phys. Rev. &D48 (93) 3768.}

\lref\bohmbook{D.~Bohm, {\it ``Quantum Theory''}, Prentice-Hall,
Englewood Cliffs, N.J. (1951), Chapter 22.}

\lref\penrose{R.~Penrose, {\it ``The Nature of Space and Time''}, Isaac
Newton Institute debate (with S.W.~Hawking), May 1994.}

\lref\penrosebook{R.~Penrose, {\it Shadows of the Mind}, Oxford
University Press, Oxford, (1994).}

\lref\mwi{ B.~DeWitt and R.N.~Graham,  eds. {\it The Many Worlds
Interpretation
of Quantum Mechanics}, Princeton University Press, Princeton (1973).}

\lref\everett{H.~Everett, \ajou Rev. Mod. Phys. &29 (57) 454.}

\lref\bell{J.S.~Bell, {\it ``Quantum Mechanics for Cosmologists''}, in
{\it Quantum Gravity 2}, eds. C.~Isham, R.~Penrose and D.~Sciama,
Clarendon Press, Oxford (1981) 611-637.}
\lref\bellthm{J.S.~Bell, \ajou Physics &1 (64) 195; \ajou
Rev. Mod. Phys. &38 (66) 447.} 
\lref\page{D.~Page, \ajou Phys. Rev. Lett. &70 (93) 4034.}

\lref\goldsteinpage{S.~Goldstein and D.~Page, \ajou Phys. Rev. Lett. 
&74 (95) 3715.}

\lref\jooszeh{E.~Joos  and H.D.~Zeh, \ajou Zeit. Phys. &B59 (85) 223.}

\lref\zeh{ H.D.~Zeh, \ajou Found. Phys. &1 (71) 69.}

\lref\zurek{W.~Zurek, \ajou Phys. Rev. &D24 (81) 1516;
\ajou Phys. Rev. &D26 (82) 1862.}

\lref\caldlegg{A.~Caldeira  and A.~Leggett, \ajou Physica &121A (83) 587.}

\lref\saunders{S.~Saunders, {\it ``The Quantum Block Universe''},
Harvard Dept. of Philosophy preprint (1992);
{\it ``Decoherence, Relative States, and Evolutionary Adaptation''},
Harvard Dept. of Philosophy preprint (1993).}

\lref\bellmwicrit{J.S.~Bell, {\it ``The Measurement Theory of Everett and
De Broglie's Pilot Wave''}
in {\it Quantum Mechanics, Determinism, Causality and Particles}, M.~Flato
et al. (eds.), D.~Reidel, Dordrecht, 1976; reprinted in
J.S.~Bell, {\it Speakable and Unspeakable
in Quantum Mechanics}, Cambridge University Press, Cambridge (1987).}

\lref\stein{H.~Stein, \ajou No\^{u}s &18 (84) 635.}

\lref\ak{A.~Kent, \ajou Int. J. Mod. Phys. &A5 (90) 1745.}

\lref\dktruth{F.~Dowker and A.~Kent, {\it ``A New `Truth'''}, in
preparation}

\lref\despagnat{B.~d'Espagnat, \ajou J. Stat. Phys. &56 (89) 747.}

\lref\samols{T.M.~Samols, {\it ``A Stochastic Model of a Quantum Field
Theory,''} Cambridge preprint DAMTP/94-39, to appear in 
{\it J. Stat. Phys.}}

\lref\sorkin{R.D.~Sorkin,
{\it ``Quantum Mechanics as Quantum Measure Theory,''} Syracuse preprint
SU-GP-93-12-1, gr-qc/9401003.}

\lref\bohm{D.~Bohm, \ajou Phys. Rev &85 (52) 166.}

\lref\grw{G.~Ghirardi, A.~Rimini and T.~Weber, \ajou Phys. Rev. &D34
(86) 470.}

\lref\gisin{N.~Gisin, \ajou Helv. Phys. Act. &62 (89) 363.}

\lref\percival{I.~Percival, \ajou Proc. Roy. Soc. Lond. Ser. A &447
(94)
189.} 

\lref\vonn{J.~von Neumann, {\it Mathematical Foundations of Quantum
Mechanics},
Princeton University Press, Princeton (1955).}

\lref\dowkerkentprl{F.~Dowker and A.~Kent, \ajou Phys. Rev. Lett. &75
(95) 3038.}

\lref\gosse{P.H.~Gosse, {\it Omphalos: an attempt to untie the geological
knot}, (1857).}

\lref\wigner{E.P.~Wigner, {\it ``Remarks on the Mind-Body Question''} in
{\it The Scientist Speculates}, I.J.~Good, ed., pp. 284-302,
Heinemann, London (1961).}

%%Start real stuff

\Title{\vbox{\baselineskip12pt\hbox{ DAMTP/94-48}\hbox{ NI 94006}
\hbox{ gr-qc/9412067}
}}
{\vbox{\centerline {On the Consistent Histories}\vskip2pt
\centerline{Approach to Quantum Mechanics}
}}{August 1994~(revised June 1995)}

\centerline{{\ticp Fay Dowker}${}^{1,2}$}
\centerline{{\ticp Adrian Kent}${}^{3}$}

\vskip.1in

\centerline{\sl ${}^1$Physics Department, University of
California-Santa Barbara,}
\centerline{\sl Santa Barbara, California 93106, U.S.A.}
\vskip.1in
\centerline{\sl ${}^2$Isaac Newton Institute for Mathematical Sciences}
\centerline{\sl 20 Clarkson Road, Cambridge CB3 0EH, U.K.}
\vskip.2in
\centerline{\sl ${}^3$Department of Applied Mathematics and
Theoretical Physics,}
\centerline{\sl University of Cambridge,}
\centerline{\sl Silver Street, Cambridge CB3 9EW, U.K.}
\bigskip

\centerline{\bf Abstract}
{We review the consistent histories formulations of quantum
mechanics developed by Griffiths, Omn\`es and Gell-Mann and Hartle,
and describe the classification of consistent sets.
We illustrate some general
features of consistent sets by a few simple lemmas and examples.
We consider various interpretations of the formalism, and
examine the new problems which arise in reconstructing the past and
predicting the future.
It is shown that Omn\`es' characterisation of true statements --- statements
which can be deduced unconditionally in his interpretation --- is
incorrect.
We examine critically Gell-Mann and Hartle's
interpretation of the formalism, and in particular their discussions
of communication, prediction and retrodiction, and conclude that their
explanation of the apparent persistence of quasiclassicality relies
on assumptions about an as yet unknown theory of experience.
Our overall conclusion is that the consistent histories approach
illustrates the need to supplement quantum mechanics by some selection
principle in order to produce a fundamental theory capable of
unconditional predictions.
}

\newsec{Introduction}
\footline={\ifnum\pageno=0 {\hfil} \else\hss\tenrm\folio\hss \fi}

Quantum theory has been so successful that it is natural to wonder whether
it can be applied to the universe as a whole.
It has long been
clear that to do any such thing we would need a new way of thinking
since, while the standard Copenhagen interpretation
works well in the laboratory, it cannot be applied to
closed systems.
Recently Griffiths, \refs{\griff,\ \grifflogic} Omn\`es
\refs{\omnes,\ \omnesreview} and Gell-Mann and
Hartle \refs{\gmhsantafe,\ \gmhone,\ \gmhtwo,\ \gmhthree,\ \gmhprd} have
indeed set out a new way of looking at quantum
mechanics and quantum field theory, in which the fundamental objects
are consistent sets of histories. So, can the consistent
histories formalism do the job required?  Could it be the correct language in
which to give an intrinsically quantum description of cosmology?  Is it
complete?  Our aim here is to examine these questions.
We shall attempt to clarify and extend the arguments in the literature,
to make explicit the premises on which they rely, and to clarify
the present status of interpretations of the consistent histories
formalism as scientific theories.  
We briefly sketch our approach in the rest of this introduction.

We should say, firstly,  
that the present article contains 
a much greater word to equation ratio 
than is usual in a scientific paper. We have found this,
unfortunately, to be 
unavoidable since, 
while some of the questions that arise in the area are technical,
the deepest controversies involve
the interpretation of the formalism and 
are not to be resolved simply by calculating.  
Many contradictory arguments have been advanced, and
we have tried comprehensively to deal with the main proposals.  
A short letter \refs{\dowkerkentprl}
illustrating the main features of the consistent 
histories formalism may be of some help to the reader.  

By a quantum cosmological theory we mean some precise,
observer-independent formulation of quantum theory (such as,
perhaps, a consistent histories formulation) together with some theory
of the boundary conditions.  Without a quantum theory
of gravity, the best one can hope for is an effective theory.
However, given a fixed spacetime with preferred timelike
directions, and given some theory about
the density matrix on some very early spacelike surface, the question
remains: what should, or could, we ask from such a theory?
It is surely too much to hope that it will predict in detail the
world we experience.  However one might at least hope to predict the
gross features: the existence and persistence of
(what appears to us to be) a classical world with very large-scale
structure; the fact that matter clumps and fields in this classical
world generally obey classical equations of motion to a very good
degree of approximation; the fact that microscopic particles, when
sporadically interacting with the classical world, follow the
probabilistic laws of Copenhagen quantum theory.

Obtaining predictions such as these from a formulation of quantum
theory is notoriously tricky.  It clearly is not good enough to show that
by some procedure
one can obtain from quantum theory a picture which gives a
satisfactory account of cosmology, if one can equally well extract by
similar procedures many other pictures which give completely
different, unsatisfactory accounts.  So we must first set out a way of
obtaining cosmological data from quantum theory.
Given any such procedure, we must then
characterise all the sets of data which it gives.
Only then can we tell whether we have a sensible and accurate quantum
theory of cosmology, a potentially useful theory with an unresolved
selection problem, or something less coherent.
The theory, in other words, is more than just the equations: it
includes an explanation of how the equations are to be interpreted.

There does seem to be a consensus among physicists that these demands
are reasonable.
Still, it can be tempting to
extract physical facts from the mathematics whenever they are
appealing, to refuse to do so whenever they would be embarrassing, and
not to enquire too closely as to whether any discernible principle
underlies these choices.
In particular, when a new approach, such as the consistent histories
formalism, is discovered, it is sometimes suggested that the only
interesting problem is extending the mathematics, and that
interpretation is just a matter of common sense.
Yet this view cannot easily be maintained, once one recalls the
intensity of the debates which led eventually to the Copenhagen
interpretation.
It seems even more naive when one considers the history of
various attempts at a many-worlds interpretation.\foot{A collection
of early papers on the subject can be found in DeWitt and
Graham's book.\refs{\mwi}
A detailed discussion of many-worlds interpretations
is beyond our present scope; several critiques can be found in the
literature.\refs{\bellmwicrit,\ \stein,\ \ak}
Briefly, our view is that although the ideas of Everett et al.
have motivated interesting work, including some of the papers we
shall discuss, no well-defined scientific theory has yet been
described in the many-worlds literature except for Bell's intentionally
pathological Everett-de Broglie-Bohm hybrid,\refs{\bell}
and that most of the ideas hinted at in earlier
many-worlds papers can more naturally be understood in the language of
consistent histories.}
And it is certainly not the view of the authors of
the consistent histories formalism, who have each set out their
interpretational ideas at some length.

We shall try, then, to be careful to specify exactly how one translates
the consistent histories formalism into physics.
Fortunately, as we describe later, there are a number of ways
to do this, so that it is
possible to use the formalism to obtain physical results.
To understand the nature of these results, we need some description
of the various consistent sets and some characterisation of the events they
each describe.
In section 2, we set out the formalism, discuss the various
consistency criteria which have been proposed, and examine equivalence
relations and maximality criteria for consistent sets.

Any realistic discussion of the general properties of consistent sets
in cosmological models is hopelessly impractical. In fact, exploratory
calculations on finite-dimensional Hilbert spaces show that,
even given a simple initial density matrix, the spaces of consistent
sets of histories become complicated very quickly; it is a formidable task
to calculate the full class of consistent sets 
of histories in dimensions greater than about three.
In section 3 we set out the general algorithm for classifying
consistent sets, and give some fragmentary results describing the
properties of the space of consistent sets.
These results, which are obtained by exact calculation of the very
simplest examples, give a limited and qualitative test of the
conclusions one can draw from naive counting arguments.
Fortunately, we need these results and counting
arguments only to
support propositions which already seem very plausible.
We also set out some elementary but useful lemmas and illustrative
examples which seem to characterise the basic mathematical facts that
lead to problems --- or, at least, unfamiliar features --- in the
various interpretations of the formalism.

In section 4 we briefly examine two of Gell-Mann and Hartle's
proposals.  
The first is the use of approximate consistency.  It is easy to set out
the conditions for a set of histories to have probabilities satisfying
the usual sum rules.  Gell-Mann and
Hartle \refs{\gmhsantafe,\ \hartleone} argue that
it is natural to consider, on an equal footing
with these consistent sets
of histories, those sets for which the probability sum rules are
slightly violated.  They point out that, if the violation is
sufficiently small, no experiment could detect the discrepancy, and
that in any case one can remove the sum rule violation by an ad hoc,
but equally undetectable, renormalisation of the probabilities.
Still, this seems a rather casual disruption of the mathematical
structure of a fundamental theory.
It also seems unnecessary.  The intuition one can glean
from calculations of consistent histories on finite-dimensional
Hilbert spaces is admittedly limited, but it
does suggest that the solution space of the exact consistency
equations is quite large enough for all theoretical purposes so far
suggested.  In particular, naive but plausible counting arguments
suggest that, in the neighbourhood of generic approximately consistent
sets of histories, an exactly consistent set can be found.  This,
we shall suggest, means that there is no need to consider approximately
consistent sets in any fundamental formulation of the theory.
Given that there are many consistent, incompatible sets of histories, it is
natural to ask whether there might be any good way of singling one out.
The second proposal of
Gell-Mann and Hartle that we comment on
is that there is some simple and natural measure
of the quasiclassicality of consistent sets, which is sharply peaked
\refs{\gmhonmeasure}.
While this goes beyond the 
framework of consistent histories, it is an attractive idea.
Finite-dimensional history calculations give no clue as to its
plausibility, but illustrate in a little more detail the
questions involved.

In section 5, we set out various possible interpretations of the
formalism, and analyse their status as scientific theories.
Our main preoccupation here is the extent to which any interpretation
allows us to make deductions about the past or present and predictions
of the future.  We explain that there are several different senses in which
a consistent histories theory can be said to account for the past, and
try to determine which particular sense is intended by particular
authors.
We then attempt to understand the extent to which
any consistent histories interpretation can account for our perception
of a single, persisting, nearly classical world.
Since one can show that, unless the
particular set of histories which describes our experiences up 
to the present has
remarkable and previously unsuggested
properties, present-day quasiclassicality in a consistent set
does not imply persisting quasiclassicality in that set, and in 
fact quasiclassicality does not persist in generic
consistent future extensions, this is clearly a non-trivial problem.
We also attempt to understand the extent to which consistent histories
interpretations allow us to make ordinary physical inferences.
Can we show, for instance, from terrestrial experiments,
that the moon is following an essentially classical orbit around the
earth, or more generally that the world around us is
approximately described by classical physics?

Here we take the interpretations case by case.
We first consider the logical interpretation of the formalism which
has been set out by Omn\`es.
Omn\`es uses the formalism to analyse the logical relations amongst various
collections of statements, and reaches an interpretation of
the formalism which is logically consistent and which, in a certain sense,
extends the Copenhagen interpretation.
There are interesting ideas in this approach.
However, we conclude that on a key point --- the identification of
true statements which, given a particular set of data, can
be deduced unambiguously --- Omn\`es' conclusions
are incorrect.  We argue that in fact there are generally
no useful true statements describing the future, and only a very
limited class of true statements about the past.
Omn\`es accepts this criticism.  Work on a revised criterion for 
``truth''
is in progress: as we shall explain, we are sympathetic to this
program, though we believe it will necessarily go beyond the
consistent histories formalism.
Meanwhile, as presently formulated, Omn\`es'
interpretation does not allow us to make any unconditional
tests of the formalism, and in particular gives no explanation for
the apparent persistence of quasiclassicality.

We next mention an interpretation set out by
Griffiths, which relies on a notion of quantum logic rather different
from that of Omn\`es.
As we explain, Griffiths' logical structure, in itself, allows no
useful notion of prediction or retrodiction.
We then set out a many-histories interpretation, inspired by Griffiths' ideas
but requiring only classical logic.
This natural-seeming interpretation is quite different
in spirit from the previous two; it allows different arguments to
be made and suggests different directions for possible extensions of
the formalism.  However, its practical implications are essentially
the same: the past is unreconstructable, and the persistence of
quasiclassicality is inexplicable.

We come next to Gell-Mann and Hartle's work on the interpretation of
the consistent histories formalism in the light of quantum cosmology.
Gell-Mann and Hartle argue that creatures
such as ourselves, whose evolution is best described quasiclassically, will
naturally perceive a persisting quasiclassical world.\refs{\gmhigus}
This is an important claim which deserves careful scrutiny: if it holds, then
the consistent histories approach, together with a theory of the boundary
conditions, constitutes a genuinely predictive scientific theory; if it
fails, then it seems very unlikely that any useful and unconditional
predictions of our future observations can be made using the existing
formalism.
We attempt to analyse Gell-Mann and Hartle's arguments and to identify
the premises on which they rely.
We conclude that the argument that quasiclassicality will appear to
persist relies crucially on assumptions about an as yet unknown
theory of experience.  We note also that, although they often write in
ordinary language about the reality of past events and the possibility
of communication, the only interpretation we can find in which
Gell-Mann and Hartle's prediction of our experience of a 
quasiclassical world can be hoped to be made entails solipsism.  
 
In the final subsection on interpretations, we describe a minimalist
interpretation in which our continuing perception of a quasiclassical
world is very clearly an unexplainable mystery. 
We argue that the Omn\`es, 
Griffiths and many histories  
interpretations can all be characterised as variants of this
``unknown set'' interpretation.

The apparent persistence of quasiclassicality is central to our
discussions.  In section 6 we look at another aspect of the question.
We show that generic cosmological theories which contain
a consistent set describing quasiclassical behaviour up till
some time do not contain a consistent
set in which quasiclassicality persists.
More precisely, we show that,
if the initial density matrix is pure and some consistent collection
of present and past facts is given, and if this collection of facts
has probability $p$ when the final density matrix is trivial then,
given any non-trivial statement about the future, then  
generic final density matrices for which the given facts are also consistent
with probability $p$ render the future statement 
inconsistent.  In other words, it is impossible to ascribe
consistency to {\it any} non-trivial statement about the future
without invoking some general theory of the final density matrix.
In section 7, we examine the case for imposing any consistency
criterion.  We point out that there are logically coherent interpretations
involving inconsistent sets, and indeed that it is not completely
impossible that the world as we see it is described by an
inconsistent set.  However, the success of
experimental science to date is a compelling argument against
this possibility.

In the concluding section, we summarise the earlier discussions and
our view of the current status of the consistent histories formalism
and its interpretations.
We concur with its authors that it is a natural approach to
quantum theory. 
However, we conclude that, whichever
interpretation one adopts, the formalism at present provides only a very
weakly predictive scientific theory.  We suggest that the formalism
ought to be augmented, and briefly discuss how this could be done.

We close this introduction with a few historical remarks.
The consistent histories formalism has evoked a good deal of interest
over the last decade, and has provoked many thoughtful
criticisms.
Griffiths' original article \refs{\griff} included a discussion of
some novel and counterintuitive features of consistent sets of histories,
and he, Omn\`es, and Gell-Mann and Hartle have since devoted much
effort to elaborating and explaining the consequences of such features.
In a thoughtful critique,\refs{\despagnat}
whose conclusions we cite later, D'Espagnat examined the early
papers of Griffiths and Omn\`es, and noted some divergence
between a language of definite physical events and
formal discussions in which these reassuringly realistic pictures
cannot quite be justified.  Griffiths \refs{\grifflogic} and
Omn\'es \refs{\omnesreview} have developed their
ideas further in response to D'Espagnat.
Among other critics, Zurek, in an illuminating discussion \refs{\zurekptp}
of decoherence-based approaches to the
problems of quantum mechanics, has emphasized that consistency does
not constrain histories enough to eliminate even very
non-classical evolutions of the state vector, and argued for
an interpretation based on the division of the universe into
subsystems.
D\"urr, Goldstein and Zangh\'\i, (DGZ) in their careful and detailed
analysis\refs{\duerretal} of Bohmian mechanics, noted the
analogy between the trajectories of Bohmian mechanics and 
Gell-Mann and Hartle's speculative program aimed at understanding
how quasiclassical domains might be characterised and investigating
their properties and the possibility that an essentially unique
domain might exist.\foot{To quote DGZ: ``they [Gell-Mann and Hartle] 
propose a program to extract from the quantum formalism a 
`quasiclassical domain of familiar experience' which, if
we understand them correctly, defines for them the basic ontology
of quantum mechanics''.  In fact, as we explain later, DGZ do not 
understand Gell-Mann and Hartle correctly on this point, 
although we agree this would be a clearer and more
natural interpretation than any Gell-Mann and Hartle offer. 
With this caveat, we recommend DGZ's remarks (Appendix of
\refs{\duerretal}, point 22) to the reader.} 
Albrecht\refs{\albrecht} discussed the relationship
between descriptions of a combined system and apparatus in terms
of consistent histories and of the Schmidt decomposition, with
the aid of a series of computational experiments.
Paz and Zurek\refs{\pazzurek} have further explored this relationship
theoretically, and shown that, when particular dynamical
approximations hold, a large class of branch-dependent
consistent histories can be identified.

It has, in short, been recognised for some time that,
in the consistent histories approach to quantum
cosmology, there are likely to be a large number of consistent sets,
and that many of these will not be
standard quasiclassical descriptions of the world.
Whether this causes scientific problems (and if so which), and whether
these problems can be dealt with by simple postulates or
interpretational arguments, are questions which
depend very much on the details.
We here try, inter alia, to characterise much more precisely the
general properties of consistent sets, to analyse carefully the
scientific problems which these properties cause, and to engage in
detail with the interpretational arguments which Omn\`es, Gell-Mann
and Hartle, and Griffiths offer in defence of the formalism.

There are, of course, interesting formulations of quantum theory 
outside the consistent histories framework --- most notably, in our
view, Bohmian mechanics\refs{\bohm} and Samols' recent stochastic model
of a relativistic quantum field theory.\refs{\samols}    
There is no doubt that these approaches solve some of the problems
afflicting current interpretations of the consistent histories
formalism.  In particular they explain the persistence of 
quasiclassicality by introducing auxiliary variables --- often,
misleadingly, called ``hidden variables'' --- which, very roughly
speaking, do the job that a good set selection criterion would do 
within the consistent histories framework.   
On the other hand these approaches raise questions (of relativistic
invariance in the case of Bohmian mechanics, and of 
generalisability and naturality) of their own.  
These questions, though very interesting, go beyond our scope, and
we shall generally avoid comparisons of the consistent histories
framework to other approaches: the interested reader is encouraged
to consult Refs. \refs{\duerretal,\ \samols,\ \cfg}. 

\newsec{Consistent Histories}

\subsec{Formalism}

It is simplest to describe the consistent histories formalism as it
applies to non-relativistic quantum mechanics, in the
Heisenberg picture, using the language of projection operators and density
matrices.
As in standard quantum mechanics, we assume that a separable
Hilbert space, $\cal{H}$, 
and hamiltonian, $H$,  are given, that hermitian operators correspond to
observables,
that the commutation relations amongst the hamiltonian and physically
interesting observables (such as position, momentum and spin) have
been specified, and
that the operators corresponding to the same observables at different times
are related by
\eqn\heis{
P (t) = \exp(i H t / \hbar) \, P(0) \, \exp( - i H t / \hbar ) \, .}
We are interested in a system
(in principle, the universe) whose initial density matrix $\rho_i$ is
given.  As usual, we require that $\Tr ( \rho_i ) = 1 $ and that
$\rho_i$ is hermitian and positive semidefinite.  

The consistent histories formalism also admits an interesting
time-symmetric generalisation of quantum mechanics,\refs{\abl} in which both
initial and final boundary conditions are specified via $\rho_i$ 
and a {\it final} density matrix
$\rho_f$ which is hermitian positive semidefinite and
normalised\foot{
With this normalisation, since $\Tr (\rho_f ) 
\ne 1$ in general, $\rho_f$ is not, strictly speaking, a density
matrix.  We shall nonetheless refer to it as such, since both the 
normalisation and the terminology are convenient.}
so that $\Tr ( \rho_i \rho_f ) = 1$.  
This is particularly interesting in
the context of quantum cosmology, removing as it does the usual
time asymmetry in the boundary conditions.\refs{\rhofhartle}
We shall often include a final density matrix, but, where the
value of $\rho_f$ is not explicitly discussed,
it can be set to $I$ without affecting the arguments.
We shall use the convention that, where non-trivial final density matrices are
excluded, the initial density matrix is called $\rho$ rather than $\rho_i$.
The initial and final density matrices give boundary conditions for the
system at times $t_i$ and $t_f$, with $t_i < t_f$; these times can generally
be set to $0$ and $\infty$.  The basic physical
events we are interested in correspond to sets of orthogonal hermitian
projections $P^{(i)}$, with
\eqn\conds{\sum_i P^{(i)} = 1 \quad {\rm{and}}\quad
P^{(i)} P^{(j)} = \delta_{ij} P^{(i)}.}
We shall want to consider the projections as attached to some particular
time $t$; where we wish to emphasize this, we shall 
write $P^{(i)}$ as $P^{(i)} (t)$.
It may be useful to visualise these as projections onto disjoint
intervals of the spectrum of some observable, for example the position
of a single particle.  For a single set of this type, with trivial
$\rho_f$, standard quantum
mechanics supplies an interpretation: specifying the set corresponds
to specifying the possible results of a particular experimental
measurement at time $t$, and --- if the experiment is carried out ---
the probability of obtaining result $i$ is
\eqn\probone{p(i) = \Tr ( P^{(i)} (t) \rho_i P^{(i)} (t) ) \,  .}
%           \Tr ( \rho_i \rho_f ) \,
%

In fact, so long as a single set of projections has somehow been
fixed, it is perfectly consistent to speak about the state of the
system without invoking the notion of measurement.  If we wish, we
can simply postulate that the projections define an extra variable
which represents the ``real'' physical description of the system,
and say that at time $t$ the system {\it is} in the range of precisely
one of the projections, and the probability is $p(i)$ that this projection is
$P^{(i)}(t)$ --- although, of course, the formalism does not imply
this interpretation.  

One needs to take more care in assigning probabilities to
{\it{histories}} --- that is, to sequences of events in time.
We now allow non-trivial $\rho_f$, and suppose we have sets of projections
$\sigma_j =\{ P^{(i)}_j; i = 1,2,\ldots , n_j \}$
with $n_j>1$ and  $j$ running from $1$ to $n$, at times $t_j$
with $t_i<t_1<\ldots<t_n<t_f$, which satisfy the conditions \conds.
Then the histories given by choosing one projection from each
$\sigma_j$ in all possible ways are an exhaustive and exclusive set of
alternatives, $\cal S$.
It seems natural to interpret a history
$\{ P^{(i_1 )}_1(t_1), \ldots ,P^{(i_n )}_n (t_n) \} $,
where $P^{(i_j )}_n (t_j) \in \sigma_j$, as corresponding to
the proposition that the system was in the range of $P^{i_1}$ at time $t_1$,
in the range of $P^{(i_2)}$ at time $t_2$, and so on.
The natural rule for the probability of this history
is\foot{This will be recognised as the formula for the
probability, calculated in the Copenhagen interpretation,
that the history is  {\it measured} in a
sequence of ideal measurements.}
\eqn\probtwo{ p(i_1 \ldots i_n) = \Tr (  P^{(i_n )}_n \ldots
P^{(i_1 )}_1 \rho_i P^{(i_1 )}_1 \ldots P^{(i_n )}_n \rho_f ) \, .}
However, one would also like the
standard probability sum rules for exclusive events
to be reflected in the mathematics in the natural way.  We shall discuss this
point further in section 6, but let us give the standard argument here.
It is that the projection operator $(P^{(a )}_r + P^{(b )}_r )(t_r)$
(where $a$ and $b$ are distinct labels in the range $1$ to $n_r$)
should correspond to the statement that the system was in the union of
the ranges of $P^{(a )}_r$ and $P^{(b )}_r$ at time $t_r$, and the
probability of a history including this event should be calculable either
directly, using this projection operator, or as the sum of the probabilities
of the two finer-grained histories in which the system was, respectively, in
the range of $P^{(a )}_r$ and in the range of $P^{(b )}_r$.
That is, for consistency we require that
\eqn\sumrules{\eqalign{
\quad\Tr (P^{(i_n )}_n \ldots
(P^{(a )}_r + P^{(b )}_r ) \ldots
P^{(i_1 )}_1 \rho_i P^{(i_1 )}_1 \ldots
(P^{(a )}_r  + P^{(b )}_r ) \ldots
P^{(i_n )}_n \rho_f) = \  \  \  \  \hfill{} \cr
\hfill{}\Tr ( P^{(i_n )}_n \ldots
P^{(a )}_r  \ldots
P^{(i_1 )}_1 \rho_i P^{(i_1 )}_1 \ldots
P^{(a )}_r \ldots
P^{(i_n )}_n \rho_f ) \, \, + \, \, \, \, \cr
\hfill{}\Tr ( P^{(i_n )}_n \ldots
P^{(b )}_r  \ldots
P^{(i_1 )}_1 \rho_i P^{(i_1 )}_1 \ldots
P^{(b )}_r \ldots P^{(i_n )}_n \rho_f ) \, \, , \, \cr}}
and similar conditions involving sums of more than two projections.
These conditions do not hold for general sets $\cal S$.
As Griffiths\refs{\griff} pointed out, equation \probtwo\
will be consistent with these probability sum rules
if, and only if, the projection
operators satisfy
\eqn\decohgriff{ {\rm Re}\,
(\Tr ( P^{(i_n )}_n \ldots P^{(i_r )}_r \ldots P^{(i_1 )}_1
                               \rho_i P^{(i_1 )}_1
\ldots P^{(i'_r )}_r \ldots P^{(i_n )}_n \rho_f )) =
\delta_{i_r i'_r } p(i_1 \ldots i_n ) \, ,}
for all $r$ and all choices of $i_1 , \ldots , i_n$ and $i'_r$.
We shall follow Gell-Mann and Hartle in imposing the stronger condition that
\eqn\decohgmh{ \Tr ( P^{(i_n )}_n \ldots P^{(i_1 )}_1
                               \rho_i P^{(j_1 )}_1
\ldots P^{(j_n )}_n \rho_f ) =
\delta_{i_1 j_1 } \ldots \delta_{i_n j_n } p(i_1 \ldots i_n ) \, .}
As Gell-Mann and Hartle stress, it is natural
to require \decohgmh\ if one
is modelling the physical decoherence mechanisms of quasiclassical
variables; it also seems to be mathematically the most natural, and
most convenient, consistency condition.
When the conditions \decohgmh\ hold, we say that the set of
histories, $\cal S$, is
consistent.\foot{The term {\it medium decoherent}, coined
by Gell-Mann and Hartle, is used in the literature to describe sets
satisfying \decohgmh.  {\it Weakly decoherent} sets satisfy only the real
part of \decohgmh.  This condition is motivated by
the path integral 
version of the non-relativistic theory, in which not only
individual projections but entire histories are required to obey
probability sum rules.  Note that Griffiths'
original consistency condition \decohgriff\
is weaker still. For more comments see \hbox{Ref. \refs{\gmhprd}.}
We prefer the condition \decohgmh\ as explained in the
text, and Griffiths' terminology because of its implication of
virtue.}
We use \decohgmh\ rather than the Griffiths conditions \decohgriff\
because our general preoccupation will be with the number of
consistent sets and the problems this causes in prediction and
retrodiction.  We look favourably on any natural
criterion which reduces the number of sets, in the hope of reducing
these problems, or at least --- since we believe that adopting \decohgriff\
would not significantly change our discussion --- in assuring
ourselves that we have done what we can.

There are other versions of the theory, for example the
path integral approach,\refs{\hartleone,\ \hartletwo,\ \gmhpathint} 
in which
a set of coarse grained histories is a partition of
the set of fine grained paths, and the generalisation to ``history
dependent projections'',\refs{\omnes,\ \gmhonbranches}
in which the set of projections at a certain time may
depend on the previous choices of projections.
Other interesting generalisations of the formalism are also being
examined.\refs{\isham}
It is clear that the non-relativistic formalism which we examine here
cannot be fundamental,
and it will be necessary eventually to investigate whether
the other proposals give so radically different a picture of quantum
cosmology that they evade the difficulties we discuss later.
However, this seems unlikely at first sight, and
since it anyway seems sensible to look first at the simplest
non-relativistic form of the theory, we shall restrict our attention in this
work to the projection operator formalism described above.

\subsec{Characterising Histories}

We turn to the question of when we are to regard two sets of histories
as different.  There clearly are equivalence relations which can
naturally be imposed on the consistent sets.\refs{\hartleone}
It would be useful to characterise such relations completely, but this
has not yet been done, and may not even be a well-defined problem.
Here we just make a few simple observations.

Let us start with the complete characterisation of a set of
histories that we have so far:
\eqn\char{
{\cal {S}} = (\rho_i, t_i , \rho_f , t_f , \{\sigma_j\}, \{t_j\})
}
with  $j=1,2,\dots, n$ and $t_i<t_1<\dots<t_n<t_f$ and the $\sigma_j$
complete sets of orthogonal projections,
$\{ P^{(i)}_j : i = 1,2,\ldots , n_j\}$,  as above. We are interested in
possible equivalence relations on the set of all
such sets of histories.
In the following discussion, we denote possible relations by $\sim$, and
take their reflexivity, symmetry and transitivity for granted.
Thus, for example, if we impose the relation $A \sim B$, we also
require $B \sim A$.

The first question to decide is whether any equivalence relations
should preserve the physical interpretation of the consistent sets we
actually use for calculations, or whether it is only the 
algebraic structure that need be preserved. 
The first seems more natural if one accepts the argument (which we
shall make later) that some selection
criterion will have to be found to explain why we use particular sets,
since one would hope that a good selection criterion would rely on
the hamiltonian and canonical variables, would assign times to the
relevant projections, and so forth.
If one intends to use the formalism while treating all sets
democratically, the second alternative is probably more sensible.

We start with the first, wherein equivalence relations 
respect the full hamiltonian structure and thus
should preserve the time coordinates in \char.
Equations \decohgmh\ are invariant under arbitrary unitary
transformations but, if we take the hamiltonian $ H $ to be a
fixed operator, then we consider consistent sets to be 
physically equivalent only if related by a hamiltonian preserving
unitary map, that is:
\eqn\unitarya{(\rho_i, \rho_f , \{\sigma_j\})
\sim(U \rho_i U^{-1}, U \rho_f U^{-1}, \{U \sigma_j U^{-1}\})
}
if and only if $U$ is a unitary operator such
that $ H = U {H} U^{-1}$.  Here
$U \sigma_j U^{-1}$ is shorthand
for the set of projections
$\{ U P^{(i)}_j U^{-1}: i = 1, \dots, n_j\}$.
We can alternatively, following Gell-Mann and Hartle,\refs{\gmhcomm}
consider equivalences between complete theories, which include the
hamiltonian; in this case we have
\eqn\unitarya{( {H} , \rho_i, \rho_f , \{\sigma_j\})
\sim( U {H} U^{-1} ,
U \rho_i U^{-1}, U \rho_f U^{-1}, \{U \sigma_j U^{-1}\}) \, .
}

On the other hand, even if one anticipates a set selection criterion,
the arguments in favour of restricting to hamiltonian-preserving equivalences
are compelling but not quite conclusive.  Many of the consistent sets
in a theory describe a rich mathematical structure, even if a hamiltonian
is not given: it may seem implausible that this structure alone could be
sufficient to characterise the quasiclassical world we see, but it is hard
to see how one could argue the point with complete certainty.

We shall take a pragmatic approach.
We chose the Gell-Mann--Hartle consistency conditions rather than the
Griffiths conditions because they were mathematically more convenient
and led to fewer consistent sets; likewise, in the calculations
we set out later we allow equivalence relations
which do not preserve the hamiltonian structure because they lead to
simpler calculations and to fewer equivalence classes.
If we restricted ourselves to relations preserving the structure, our
later arguments and illustrative calculations describing the multiplicity
of equivalence classes would only be strengthened.

So, we first assume that the only relevance of the
times $\{t_j\}$ and $\{ t_i ,
t_f \}$ is their ordering. Thus
\eqn\difftimes{
(\rho_i, t_i , \rho_f , t_f , \{\sigma_j\}, \{t_j\}) \sim
(\rho_i, t'_i , \rho_f , t'_f , \{\sigma_j\}, \{t'_j\})
}
if $t'_i<t'_1<\dots<t'_n<t'_f$, and we henceforth drop the time labels.

Secondly, we impose the relation
\eqn\unitary{(\rho_i, \rho_f , \{\sigma_j\})
\sim(U \rho_i U^{-1}, U \rho_f U^{-1}, \{U \sigma_j U^{-1}\})
}
where $U$ is any unitary operator on the Hilbert space.
(In fact, we can take $U \in SU({\cal H})$, since both sides of \unitary\
are obviously equal for $U = \exp ( i \theta) I$.)

With these relations, although the
members of an equivalence class are essentially indistinguishable as
sets of projection operators, some may be much more easily related to
simple physical observables than others.
Given a consistent set, one would have to search through its equivalence
class to see whether (for example) it could be represented in terms of
projections describing coarse grained particle trajectories.

A careful and detailed discussion is given in
Gell-Mann and Hartle's recent elegant
treatment of equivalence relations within the
formalism,\refs{\gmhcomm} which is intended to
supersede their earlier discussions.\refs{\gmhsantafe, \ \hartleequiv}
Gell-Mann and Hartle consider equivalence relations in the absence of
any hypothetical selection mechanism; their new proposals seem very
natural, and greatly refine those we have described above.
Apart from this paragraph and occasional footnotes, we have not amended
the discussion of this section in the light of the new treatment,
so as to conform with
our general approach in this and the next section, in which we simplify
arguments and calculations
by neglecting the dynamical aspects of the formalism where possible.

There are further relations which one might consider.
For example, it might seem sensible to assign no significance to the
time-ordering of
commuting sets of projection operators if no other projections intervene.
Thus, if $\sigma_k$ and $\sigma_{k+1}$ are pairwise commuting sets of
projections, and if we define $\sigma'_k$ to be the set of non-zero projections
amongst the operators
\eqn\products{
\{ P_k^{(i_k)}P_{k+1}^{(i_{k+1})} \, : \, i_k = 1, \dots, n_k, \
i_{k+1} = 1, \dots, n_{k+1} \} \, , }
one could impose the following relation:
\eqn\grain{(\rho_i, \rho_f , \{\sigma_j\})
\sim(\rho_i, \rho_f ,\{\sigma'_{j'} \})
}
where $n = n'+1$ and
\eqn\coarse{\eqalign{
\sigma_j &= \sigma'_j, \quad j = 1,\dots, k-1\cr
\sigma_j &= \sigma'_{j-1}\quad j = k+2, \dots, n' \, . \cr
}}
However, since it is less obvious that one wants all the relations implied from
this by symmetry and transitivity, we shall not impose this last relation.

We have no compelling principle to
determine exactly which equivalences should be imposed, and
quite possibly several others may eventually be adopted:
the relation of
trivial extension, discussed in the next section, is another strong candidate.
However, we shall adopt only the particularly simple relations
\difftimes\ and \unitary.  If new equivalences are suggested,
it may eventually be necessary to investigate whether they substantially
affect our arguments below; however, it seems unlikely that they
will.  Even if all sets of histories with the same collection of
non-zero probabilities were made equivalent --- and this is surely an
upper bound on what might conceivably be sensible --- we do not
believe that it would make any qualitative difference to our
arguments in practical applications.

\subsec{Other Conditions}

It is useful to have a condition of maximality when characterising
consistent sets since otherwise one has to consider all subsets of
a consistent set on an equal footing.
No doubt it is not essential,
but we adopt it in line with our general strategy of accepting
plausible rules which reduce the number of sets.
The condition we shall define here differs from that
given by Gell-Mann and Hartle \gmhsantafe\
and is tailored to our purpose of
characterising the sets.

We first need some preliminary definitions.
We say the time-ordered set
\eqn\extend{
\S' = (\rho_i , \rho_f ,\{\sigma_1,\dots, \sigma_k, \tau, \sigma_{k+1},\dots,
\sigma_n\})}
is a {\it consistent extension} of a consistent set of histories
$\S = (\rho_i, \rho_f, \{\sigma_1 , \ldots , \sigma_n \})$ by
the set of projections $\tau = \{Q^i : i = 1, \dots, m\}$ if $\tau$
satisfies \conds\ and $\S'$ is itself 
consistent.\foot{
Though for simplicity of notation we use sets of finite length, 
this and later definitions should be understood as extending to
infinite length sets in the obvious way.} 
We say the consistent extension $\S'$ is {\it trivial} if, for
each history
$\{ P^{(i_1)}_1,\dots, P^{(i_k)}_k, P^{(i_{k+1})}_{k+1}, \dots, P^{(i_n)}_n\}$
in $\S$, at most one of the extended histories
$\{ P^{(i_1)}_1,\dots, P^{(i_k)}_k, Q^i, P^{(i_{k+1})}_{k+1},
\dots, P^{(i_n)}_n \}$ has non-zero probability.
We call $\S' = (\rho_i , \rho_f ,\{\sigma_1,\dots, \sigma_{k-1}, \sigma'_{k},
\sigma_{k+1}, \dots,
\sigma_n\} )$ a {\it consistent
refinement} of the consistent set
$\S$ if $\S'$ is consistent and the projective decomposition
$\sigma'_{k} = \{ {{P'}_{k}^{(1)}} , \ldots , {{P'}_k^{(n'_k)}} \}$ is
a refinement of
$\sigma_k = \{ P_{k}^{(1)} , \ldots , P_k^{(n_k )} \}$, by which we mean that
$n'_k > n_k$ and each $P_k^{(i)}$ can be written as the sum of one or more of
the ${{P'}_k^{(j)}}$; we define {\it trivial} consistent
refinement in the same way
as trivial consistent extension.
We use the term {\it (trivial) consistent fine graining} to mean either a
(trivial) consistent refinement or a (trivial) consistent
extension. 
We extend these definitions by taking (trivial)
consistent refinement, (trivial) consistent
extension and (trivial) consistent fine graining to be transitive relations.
\foot{Gell-Mann and Hartle
use the term fine graining to cover both the notions of extension and
refinement defined in the text, without any assumption of
consistency.\gmhsantafe}
We say a consistent set $\S$ is {\it maximally extended} if it has no
non-trivial consistent extension.
Finally, we say a consistent set of histories $\S$ is {\it fully fine grained}
if it is maximally extended, has no consistent refinement, and is not
itself a trivial extension of any consistent
set.  We follow Gell-Mann and Hartle in defining a {\it maximally
refined} consistent set to be one with no (trivial or non-trivial)
consistent fine grainings.\foot{Gell-Mann and Hartle also use the
term {\it maximal} for maximally refined sets,\refs{\gmhsantafe} and
define the related notion of a {\it full}
consistent \hbox{set.\refs{\gmhtwo,\ \gmhprd}}}

At first blush, it seems natural to define a further equivalence relation by
setting $\S \sim \S'$ if $\S'$ is a trivial consistent extension or
trivial consistent refinement of $\S$.
We shall not do so because, while in some interpretations there is a
case for considering
$\S$ and $\S'$ as representatives of the same fundamental object, it
is not clear to us
that all the sets related (via transitivity) by a chain of such relations
should also be taken as equivalent.
The equivalence classes of fully fine grained consistent sets of
histories --- which for brevity
we shall call the {\it fundamental consistent sets} ---
could reasonably be taken as the fundamental objects in any given
cosmological theory, although one needs to flesh them
out by trivial extensions when describing quasiclassical domains or
the behaviour of IGUSes.
We turn now to the problem of characterising the fundamental
consistent sets.

\newsec{Properties of consistent sets}

Since physics is described by (at least) the fundamental consistent sets of
histories, we need to know how many of these sets there will be, and what
properties they are likely to have, in typical cosmological theories.
As no demonstrably good quantum cosmological theory exists, and as
it is clear that the relevant calculations would be extremely complicated
in any cosmological theory, this seems a hopeless task at present.
However, we can find a few hints about the general features of
consistent sets by looking at finite-dimensional Hilbert spaces.
Of course, it is possible that these calculations are misleading:
conceivably, there is some good cosmological theory in which the
variety of consistent sets has quite different properties from those
suggested by finite-dimensional intuition.  However, as far as we are aware,
no one has suggested that the true cosmological theory should have this
surprising feature.  It seems
reasonable to assume that finite-dimensional intuition is correct,
until a good counterargument is produced.

In this section, we generally take the Hilbert space, ${\cal{H}}$,
to be of finite dimension $n$, although we shall allow ${\cal{H}}$ to
be infinite-dimensional and separable if the same proof covers both
cases.
We generally omit the final density matrices $\rho_f$, which are taken
to be $1$ unless otherwise stated.

\subsec{Classification}

The basic objects in the formalism are the
projective decompositions of the identity
$\sigma_j =\{ P^{(i)}_j :  i = 1,2,\ldots , n_j\}$, where the
$P^{(i)}_j$ are orthogonal hermitian projections, so that
\eqn\orthog{\sum_i P^{(i)}_j = 1 \quad {\rm{and}}\quad
P^{(i)}_j P_j^{(i')} = \delta_{i i'} P^{(i)}_j.}
These decompositions are parametrised by:\medskip\noindent
(i) the set of ranks
\eqn\dims{
\{ r^{(1)}_j , r^{(2)}_j , \ldots , r^{(n_j )}_{j} \}}
of the projection operators, where
$n = \sum_{i=1}^{n_j} r^{(i)}_j$
and we take $r^{(1)}_j \geq r^{(2)}_j \geq \ldots \geq r^{(n_j )}_{j}$,
and\medskip\noindent
(ii) (up to discrete symmetries) the generalised grassmannian manifold
\eqn\manifold{ G(n ; r^{(1)}_j , r^{(2)}_j , \ldots , r^{(n_j )}_{j} ) =
{{ U(n) } \over {( U( r^{(1)}_j )
\times U( r^{(2)}_j ) \times \ldots \times U( r^{(n_j )}_{j} ) )}} \, .}
More precisely, if the first $k_1$, the next $k_2$, \dots , and the last
$k_l$ of the $r^{(i)}_j$, ordered as above, are equal
--- where the $k_j \geq 1$, and so $\sum_{j=1}^{l} k_j = n_j$ --- then the
parametrisation manifold is actually
\eqn\manifoldtrue{ G(n ; r^{(1)}_j , r^{(2)}_j , \ldots , r^{(n_j )}_{j} )
/ {( S_{k_1} \times \ldots \times S_{k_l} ) \, , }}
where $S_k$ is the group of permutations of $k$ elements.

It is easy to use this parametrisation in explicit calculations: we can
define projections $\{ P^{(1)} , \ldots , P^{(n_j )} \}$
of ranks $\{ r^{(1)} , r^{(2)} , \ldots , r^{(n_j )} \}$ by choosing an
orthonormal basis of vectors $\{ x_1 , \ldots , x_n \}$, so that
\eqn\projns{
P^{(1)} = \sum_{i=1}^{r_1} x_i (x_i )^{\dagger} \, , \quad
P^{(2)} = \sum_{i=r_1 +1}^{r_1 + r_2 } x_i (x_i )^{\dagger} \, ,
}
and so on.  The redundancies in this parametrisation correspond to the actions
of $ U( r^{(1)} ) \times U( r^{(2)} ) \times \ldots \times U( r^{(n_j )})$ and
$ S_{k_1} \times \ldots \times S_{k_l} $.  They can be eliminated, and
the equivalence relations \unitary\ and \grain\ can be imposed, at any
convenient stage of the calculation.

Thus, in principle, we can simply fix the form of the initial density matrix,
fix the ranks of the projection operators in the type of consistent set we
wish to classify, and then impose the consistency conditions \decohgmh.
These define (real) algebraic curves in the
generalised grassmannians \manifold; their intersection is an algebraic
variety whose generating polynomials can be obtained by the usual reduction
methods.
For example, one finds that in two dimensions:\medskip\noindent
\medskip\noindent
{\it{Example 1}} \qquad
If $\rho = \left(\matrix{p & 0 \cr 0 & 1-p}\right) \, ,$
with $p \ne 0,\ 1/2,\ {\rm or}\ 1$, and we set
\eqn\pz{
P(z) = {1 \over { 1 + z z^*  }}
\left(\matrix{1 & z \cr z^* & z z^*}\right) \, ,}
then the only consistent sets of length two are of the form
\eqn\fcsone{
P^{(1)}_1 = P(0) \, , \ P_1^{(2)} =  P(\infty) \, , \
P^{(1)}_2 =  P(z) \, , \ P^{(2)}_2 = P( - 1/z^* ) \, ,}
where $z$ is any non-zero complex number.  These sets are equivalent
(by \unitary) to sets with $z$ on the positive real axis, and so the
fundamental consistent sets are of the form \fcsone\ with $z$ 
positive real.
\medskip\noindent
\medskip\noindent
%\vskip 5pt
{\it{Example 2}} \qquad If $\rho =
\left(\matrix{1/2 & 0 \cr 0 & 1/2}\right) \, ,$ then the only length two
consistent sets are of the form:
\eqn\fcstwo{P^{(1)}_1 = P(z_1 ) \, , \  P^{(2)}_1 = P( - 1/{z_1^*} ) \, , \
P^{(1)}_2 = P(z_2 ) \, , \  P^{(2)}_2 = P( - 1/{z_2^* }) \, ,}
where $z_1$ and $z_2$ are any complex numbers.  Imposing the equivalence
relation \unitary, we again find that the fundamental consistent sets take
the form \fcsone, with $z$ positive real.
\medskip\noindent
More interestingly, in three dimensions:\medskip\noindent
\medskip\noindent
{\it{Example 3}} \qquad if $\rho =
\left(\matrix{1 & 0 & 0\cr 0 & 0 & 0\cr 0 & 0 & 0}\right) \, ,$ and we set
\eqn\pztwo{
P(z_1 , z_2 ) = {1 \over { 1 + z_1 z_1^* + z_2 z_2^* }}
\left(\matrix{1 & z_1 & z_2 \cr z_1^* & z_1^* z_1 & z_1^* z_2 \cr
z_2^* & z_2^* z_1 & z_2^* z_2 }\right) \, ,}
then the equivalence classes of consistent sets of length two,
involving binary projective decompositions, are
represented by:\medskip\noindent
\eqn\fcsthree{P^{(1)}_1 = P(0 , x_{12} ) \, , \hfill
P_1^{(2)} = I - P^{(1)}_1 \, , \hfill
P^{(1)}_2 = P(x_{21} , x_{22} + i y_{22} ) \, , \hfill
P_2^{(2)} = I - P^{(1)}_2 \, ,
}
where $x_{12}, x_{21}, x_{22}, y_{22}$ are real numbers such that:
\eqn\fcsthreeconds{\eqalign{
 &~ (i) \qquad x_{12} \, y_{22} = 0
%{\rm~and~}
\cr
 & (ii) \qquad {\rm~either~} \quad 1 + x_{12} \, x_{22} = 0
\quad {\rm~or~} \quad x_{12} \, ( x_{12} - x_{22} ) = 0  \, .\hfill \cr}}
That is, the solution space is a two-dimensional real algebraic variety in the
parameter space.\foot{
We note for use in Example 4 that (i) and (ii) 
are necessary and sufficient conditions for the sets
to be consistent given that the parameters are real.} 
This simple example illustrates the general
phenomenon.
Already here, though, a calculation of the equivalence classes of
fundamental consistent sets would be rather complicated,
and for general sets in higher dimensions the algebraic equations are,
practically speaking, intractable.
Fortunately, we are interested in
the qualitative features of the solution spaces in general, rather than the
precise details of particular cases.  What one expects, naively, is a
solution space of dimension given by the number of parameters
minus the number of independent consistency equations.
Now $G(n ; r^{(1)}_j , r^{(2)}_j , \ldots , r^{(n_j )}_{j} )$ has real
dimension
$n^2 - \sum_{i=1}^{n_j} (r^{(i)}_j )^2 $, while
a set $\sigma_1 , \ldots , \sigma_k$ of projective decompositions of length
$n_1, \ldots , n_k$ gives
rise to no more than $ n_1 \ldots n_{k-1} n_k \, ( n_1 \ldots n_{k-1} \, -1 ) $
real equations.  The equivalence relation \unitary\ accounts for not more than
$( n^2 - 1)$ real parameters, depending on the form of $\rho$.
This suggests a solution space of dimension no smaller than
\eqn\dimone{
\prod_{j=1}^{k} \left( n^2 - \sum_{i=1}^{n_j} (r^{(i)}_j )^2 \right) -
 n_k  \left( ( \prod_{j=1}^{k-1} n_j ) - 1 \right)  \prod_{j=1}^{k-1} n_j -
( n^2 - 1) \, .
}

For example, if we restrict ourselves to considering projective decompositions
of a $2N$-dimensional space into $N$-dimensional subspaces and consider
sets of $k$ such decompositions, the estimate \dimone\ gives
\eqn\estone{(2 N^2 )^k - 2^{k} (2^{k-1} - 1 ) - 4 N^2  \sim 2^{k} N^{2 k} }
parameters.
This suggests that there are solutions for arbitrarily large $k$.

Whether these naive calculations are generally a good guide is open to doubt,
but it is certainly possible to find inequivalent sets of
solutions for arbitrarily large $k$.  For example, consider a
space of dimension $N \geq 4$ with an orthonormal basis
$\{ e_1, \ldots, e_{N} \}$,
let $\rho = \ket{e_1} \bra{e_1}$, and let
\eqn\infinite{\eqalign{
P^{(1)}_i = \ket{e_1} \bra{e_1} + Q_1 \, , \qquad &
P^{(2)}_i = 1 - P^{(1)}_i \quad {\rm~for~} i {\rm~odd} \, , \cr
P^{(1)}_i = \ket{e_2} \bra{e_2} + Q_2 \, , \qquad &
P^{(2)}_i = 1 - P^{(1)}_i \quad {\rm~for~} i {\rm~even} \, , \cr}}
where $Q_1$ and $Q_2$ are non-commuting projections on the subspace
spanned by $e_3 , \ldots, e_{N}$.  Then, writing
$\sigma_i = \{ P^{(1)}_i , P^{(2)}_i \}$, the set
$\S_r = (\rho, \{\sigma_1 , \sigma_2 , \ldots , \sigma_r \})$
is consistent for arbitrarily large $r$, and is not equivalent to any shorter
consistent set.  The same is true, even more trivially, of 
$\S'_r = (\rho, \{ \sigma_1 , \sigma_1 , \ldots , \sigma_1 \})$, the 
$\sigma_1$ being repeated $r$ times.  

\subsec{Lemmas and examples}

As the last examples illustrate, the full space of equivalence
classes of consistent sets is not very interesting.
The sets $\S_r$ and $\S'_r$ for $r \geq 2$ are trivial extensions 
of $\S_1$ and $\S'_1$ respectively. 
We are really interested in fully fine grained sets, and
these have qualitatively different properties:\medskip\noindent
\medskip\noindent
{\it{Lemma 1}} \qquad
Let $\S = ( \rho , \{ \sigma_1 , \ldots , \sigma_k \} )$ be a
consistent set that is not a trivial extension of any consistent set,
defined on a space $\cal{H}$ of dimension $n$, with initial
density matrix $\rho$ of rank $r$.
Then the length $k$ of $\S$ obeys $k \leq r n   $.
(In particular, (i) this holds for any fully fine grained consistent
set, and (ii) if $\rho$ is pure then $k \leq n$.)\medskip\noindent
\medskip\noindent
{\it{Proof}} \qquad Let
$\rho = \sum_{i=1}^r p_i \ket{ \psi_i } \bra{ \psi_i }$, where the
$p_i$ are positive, $\sum_{i=1}^r p_i = 1$, and the $\ket{ \psi_i }$ are
the first $r$ elements of an orthonormal basis
$\ket{\psi_1} , \ldots , \ket{\psi_n}$ of $\cal{H}$;
\hbox{let $\sigma_j = \{ P^{(i)}_j : 1 \leq i \leq n_j 
\}$.}
We write $\rho^{1/2} = \sum_{i=1}^r p_i^{1/2} \ket{ \psi_i } 
\bra{ \psi_i }$ and define
the
operators $A_{ij}$ (for $i,j$ running from $1$ to $n$) on $\cal{H}$ by
$A_{ij} \ket{ \psi_k } = \delta_{jk} \ket{ \psi_i } \,$.\medskip\noindent
     The matrix $\rho$ defines a
hermitian form
$\inprod{A}{B} = \Tr \, ( A^{\dagger} \rho B ) $
on the operators on $\cal{H}$.  The form's null space $N$ is spanned by the
operators $A_{ij}$ for $r+1 \leq i \leq n$ and $1 \leq j \leq n$, and
there is a natural isomorphism
\eqn\ismorph{O' = {\rm span}
\{ A_{ij} : 1 \leq i \leq r, 1 \leq j \leq n \} 
\equiv {\rm Op (\cal{H} )} /N  \, ,}
given by $A_{ij} \leftrightarrow A_{ij} + N$.
We have a natural positive definite hermitian form, which we denote
by $\left( \, , \, \right)$, on $O'$, given by $\left( A \, , \, B \right) =
\Tr(A^{\dagger} B)$.
Since $\S$ is consistent,
\eqn\sorthog{
\tr (P^{i_k}_k \ldots P^{i_1}_1 \rho P^{i'_1}_1 \ldots P^{i'_k}_k ) =
p( i_1 , \ldots , i_k ) \delta_{i_1 i'_1} \ldots \delta_{i_k i'_k} \, ,}
so that the operators $\rho^{1/2} P^{i_1}_1 \ldots P^{i_k}_k $ are orthogonal
with respect to $\left( \, , \, \right)$.
Thus there can be no more than $r n$ histories with non-zero probability.

Now by hypothesis each of the $\sigma_j$
for $j=1,2, \ldots k$ defines a non-trivial extension of the consistent subset
$\hat{\S_j} = ( \rho , \{ \sigma_1 , \ldots ,
\hat{\sigma_j}, \ldots , \sigma_k \} )$ obtained by its deletion.
A fortiori, each of the $\sigma_j$ for $j \geq 2$
defines a non-trivial extension of the
subset $\S_j = ( \rho , \{ \sigma_1 , \ldots , \sigma_{j-1} \} )$,
so that we can obtain $\S$ as a series of such non-trivial extensions,
starting from the length one set $( \rho , \{ \sigma_1 \} )$.
This means that for each $j$ from $1$ to $k-1$,
there is some set of indices
$i'_1 , \ldots , i'_j $ such that there are two
projections $P^{i'_{j+1}}_{j+1}$ and
$P^{i''_{j+1}}_{j+1}$ in $\sigma_{j+1}$ with
\eqn\nonzero{
p( i'_1 , \ldots , i'_j , i'_{j+1} ) \ne 0 \ne
p( i'_1 , \ldots , i'_j , i''_{j+1} )  \, . }
Since each non-trivial extension thus increases the total
number of histories with non-zero probability, the lemma follows. 
\medskip\noindent
\medskip\noindent
Note also that, when $\rho$ is pure, it follows 
that a set is maximally
extended if and only if it has $n$ histories of non-zero probability. 
Another simple result in the pure $\rho$ case, which
we shall use in section 5, is:
\medskip\noindent
\medskip\noindent
{\it{Lemma 2}} \qquad Let $\S$ be a set of consistent histories that is not
maximally extended, with a pure initial state $\rho$, and let ${\cal
H}$ be either finite-dimensional or separable.
Then there exists a continuous family of non-trivial extensions
for each history in $\S$ with non-zero probability.
\medskip\noindent
\medskip\noindent
{\it{Proof}} \qquad
Let $\rho = \ket{\psi} \bra{\psi}$
and $\S= (\rho, \{\sigma_1, \ldots , \sigma_l \})$.
For
simplicity of notation, take ${\cal H}$ to be finite-dimensional. and let
$n$ be its dimension: the proof can trivially be rewritten for
separable ${\cal H}$.
Consider the subset of histories with non-zero probability, which we
can label by products of projections $S_j$ for $j = 1,2, \ldots , r$,
where each $S_j$ is an ordered product, taking one
projection from each of the $\sigma_i$.
We define the states
$\ket{h_j} = S_j \ket{\psi} /  \norm{ S_j \ket{\psi} }$.
The consistency condition means that the $\ket{h_j}$ are
orthogonal.
Since $\S$ is not maximally extended, $r < n$: and we can extend the set
of $\ket{h_j}$ to an orthonormal basis by adding states
$\{ \ket{ e_k } : k = r+1, \ldots , n \}$.
Choosing states $\ket{h_j}$ and $\ket{e_k}$, and
complex numbers $z_1 ,z_2$ with $z_1^* z_1 + z_2^* z_2 =1$, we can
define two orthogonal states
\eqn\orthog{\ket{\phi_+} = z_1 \ket{h_j} + z_2^* \ket{e_k},\quad
\ket{\phi_-} = z_2 \ket{h_j} - z_1^* \ket{e_k} \, . }
Define projections
\eqn\projects{P_\pm = \ket{\phi_{\pm}}\bra{\phi_{\pm}},
\quad P_3 = 1 - P_+ - P_-.}
which satisfy \conds, and let $\tau = \{ P_+, P_-, P_3\}$. Then
$\S' = (\rho, \{\sigma_1 , \ldots , \sigma_l , \tau \})$ is a non-trivial
consistent extension of $\S$. The histories in $\S'$ with non-zero
probabilities correspond to the states
$P_\pm\ket{h_j}$ and the $\ket{h_i}$ for
$i  j$.
Since there is a consistent extension for every choice of $z_1$
and $z_2$, the lemma holds.
\medskip\noindent
\medskip\noindent
Note also that these extensions are generically incompatible. 
The basic point here --- that, with a pure density matrix,
one can produce many consistent sets using simple vector space algebra
--- was first remarked on by Gell-Mann and
Hartle.\refs{\gmhtriv}
One immediate consequence is worth noting:\medskip\noindent
\medskip\noindent
{\it {Lemma 3}} \qquad
If the Hilbert space is infinite-dimensional and the initial
state is pure, then there exist consistent sets $\S$ which are of infinite
length and are non-trivial extensions of all their subsets.
\medskip\noindent
\medskip\noindent
{\it{Proof}} \qquad We can explicitly build such sets by repeatedly applying
the construction of the last proof in the obvious way.
\medskip\noindent
\medskip\noindent
The above results rely on the assumption that the final density matrix
is fixed to be $I$.  Taking $\rho_f$ as a variable 
changes the picture:
\medskip\noindent
\medskip\noindent
{\it {Lemma 4}} \qquad
Let $\S =
( \ket{\psi} \bra{\psi} , \{\sigma_1 , \ldots , \sigma_l \})$
be as in Lemma 2, with ${\cal H}$ finite-dimensional, and let
$\S' = ( \ket{\psi} \bra{\psi} , \{\sigma_1 , \ldots ,
\sigma_l ,\sigma_{l+1} \})$ be a non-trivial consistent extension.
Then there exists some final 
density matrix $\rho_f$ such that:\medskip\noindent
(i) $\Tr ( \ket{\psi} \bra{\psi} \rho_f ) = 1$ and $\S_f =
( \ket{\psi} \bra{\psi} , \rho_f , \{\sigma_1 , \ldots , \sigma_l \})$
is consistent\medskip\noindent
(ii) The histories of $\S_f$ have the same probabilities as
those of $\S$.\medskip\noindent
(iii) The extension $\S'_f = ( \ket{\psi} \bra{\psi} , \rho_f ,
\{\sigma_1 , \ldots , \sigma_l ,\sigma_{l+1} \})$ is 
inconsistent.\medskip\noindent
Moreover, a generic $\rho_f$ satisfying (i) and (ii) 
also satisfies (iii). 
\medskip\noindent
\medskip\noindent
{\it{Proof}} \qquad It is enough to prove the result in the case where
$\sigma_{l+1} = \{ P , 1- P \}$.  Define $r,n,$ the
non-zero history states $\ket{ h_i }$, and the
orthogonal complement basis states $\ket{e_i}$ as in the proof of Lemma 2.
Since $\S'$ is a non-trivial extension of $\S$, we can assume without loss
of generality that $P \ket{h_1}$ does not lie in the subspace spanned by
the $\ket{h_j}$.  By the consistency of $\S'$, we have
\eqn\lemmaiva{P \ket{h_1} = a_1 \, \ket{h_1}
+ \sum_{i=r+1}^n a_i \, \ket{e_i} \,, }
with $a_1 \ne 0$ and not all the $a_i = 0$, $i = r+1,\dots,n$.
Now a density matrix $\rho_f = (\rho_f )_{ij}$ satisfies
conditions (i) and (ii) provided that $(\rho_{f} )_{ij} = \delta_{ij}$ for
$1 \leq i,j \leq r$.  If the set $\S'_f$ is consistent, then in particular
\eqn\lemmaivb{
\Tr ( \rho_f P \ket{h_1} \bra{h_1} ( 1 - P ) ) = 0 \, , }
which implies that
\eqn\lemmaivc{
a_1 (1 - a_1^* ) + \sum_{i=r+1}^n ((1 - a_1^* ) a_i (\rho_f )_{1i}
- a_1 a_i^* (\rho_f )_{i1} )
+ \sum_{i,j=r+1}^n a_i a^*_j (\rho_f )_{ij} = 0 \, .}
This is not true for generic hermitian positive semidefinite $\rho_f$ unless
all the $a_i$ (for $i = r+1,\dots,n$) are zero, which contradicts 
our earlier assumption.
\medskip\noindent
\medskip\noindent
Returning to the case where $\rho_f$ is trivial, we can sharpen
the non-uniqueness statement of Lemma 2:
\medskip\noindent
\medskip\noindent
{\it{Lemma 5}} \qquad Let $\S =
( \ket{\psi} \bra{\psi} , \{\sigma_1 , \ldots , \sigma_l \})$
be as in Lemma 2, with $\cal{H}$ finite-dimensional or separable.
Then there is no projective
decomposition $\sigma_{l+1}$ such that:\medskip\noindent
(i) $\S' = (\rho, \{\sigma_1 , \ldots , \sigma_l , \sigma_{l+1} \})$
is a consistent extension of $\S$\medskip\noindent
and (ii) any consistent extension
$\S'' = (\rho, \{\sigma_1 , \ldots , \sigma_l ,
\tau \})$ of $\S$ has a consistent extension
$(\rho, \{\sigma_1 , \ldots , \sigma_l ,
\tau, \sigma_{l+1} \})$.
\medskip\noindent
\medskip\noindent
{\it{Proof}} \qquad Suppose otherwise.  We first show
that $\sigma_{l+1}$ must be a trivial extension of $\S$.

Suppose $\sigma_{l+1}$ is a  binary projective
decomposition, $\sigma_{l+1} = \{ P , 1-P \}$.
As in Lemma 2, we can define the orthonormal set of history states
$\{\ket{h_j} : j = 1,2, \ldots, r \}$ belonging to $\S$, and
extend it to an orthonormal basis
with the states $\{ \ket{ e_k } : k = r+1, \ldots , n \}$, where if
$\cal{H}$ is separable we set $n = \infty$.  We know that
$r<n$ since $\S$ is not maximally extended; we shall assume for the moment
that $r>1$.  We pick states $\ket{h_j}$ and $\ket{e_k}$ and
complex numbers $z_1 ,z_2$ as above, and define
$\ket{ \phi_{\pm}}$ and the projections $P_{\pm}, P_3$ by equations
\orthog\ and \projects.
Since $\sigma_{l+1}$ itself defines a consistent extension, we have that
\eqn\lta{ \bra{h_a } P \ket{h_b } = 0 {\rm~for~} 1 \leq a < b \leq r \, . }
Condition (ii) implies that
\eqn\ltb{ \bra{h_a } P_{\pm} P P_3 \ket{h_b } = 0
{\rm~for~} 1 \leq a,b \leq r \, . }
But this is true for any choice of $\ket{h_j}$ and $\ket{e_k}$, which means
that, by considering all combinations such that $b \ne j$ and $a=j$,
we obtain
\eqn\ltc{ \bra{ e_k } P \ket{ h_b} = 0 {\rm~for~} 1 \leq b \leq r
{\rm~and~} r+1 \leq k \leq n \, .}
Equation \ltc\ means that
\eqn\ltspan{ {\rm span} ( \{ P \ket{h_a } \} )
     \subseteq {\rm span} ( \{ \ket{h_a } \} ) \, ,}
and the same is true for $(1-P)$.  Hence $\{ P , 1-P \}$ defines
a trivial extension \hbox{of $\S$.}
If $r=1$, we can use
\eqn\ltd{ \bra{h_1 } P_{+} P P_- \ket{h_1 } = 0 }
to show that
\eqn\lte{ \bra{ e_k } P \ket{ h_1} = 0 {\rm~for~} 2 \leq k \leq n \, ,}
and again $\{ P , 1-P \}$ must be a trivial extension.
Finally, if $\sigma_{l+1}$ is a general projective decomposition
$\{ Q_1 , Q_2 , \ldots , Q_q \}$, then we can apply the above arguments to
the decompositions $\{ Q_p , 1 - Q_p \}$ (for $p = 1,2 , \ldots, q$), which
have the same consistency properties as $\{ P, 1-P \}$.
Hence \ltspan\ holds for each of the $Q_p$, and once again $\sigma_{l+1}$
trivially extends $\S$.

So, in all cases, we have that $\sigma_{l+1}$ is a trivial extension,
and without loss of generality can be taken to be binary.
We now show that this leads to a contradiction.
We have that
$\sigma_{l+1} = \{ P , 1 - P \}$
is a trivial extension of $\S$ and can be
made to any consistent extension of $\S$.
Choose the states $\{\ket{e_{r+1}}, \dots, \ket{e_n}\}$
to be eigenstates of $P$ and $(1- P)$.
Choose $a\in \{1,2,\dots,r\}$
and $k \in\{r+1, \dots, n\}$ such that
$P \ket{h_a} = \ket{h_a}$ and $(1-P)\ket{e_{k}} = \ket{e_{k}}$; we may
do this without loss of generality, interchanging $P$ and $(1-P)$ if
necessary.
Construct $\ket{e_+} = {1\over{\sqrt{2}}}(\ket{h_a} + \ket{e_{k}})$ and
let $Q = \ket{e_+}\bra{e_+}$. Then $\{Q, 1-Q\}$ consistently extends
$\S$, but $(\ket{\psi} \bra{\psi} , \{\sigma_1 , \ldots , \sigma_l,
\{Q, 1-Q\}, \{P, 1-P\} \})$ is inconsistent. This completes the proof.
\medskip\noindent
\medskip\noindent
We can also use this argument to find a useful property of trivial
extensions:
\medskip\noindent
\medskip\noindent
{\it{Lemma 6}} \qquad
Let $\S' = (\rho, \{ s_1 , \ldots , s_l , t \})$ be a trivial consistent
extension of the consistent set
$\S = (\rho, \{ s_1 , \ldots , s_l \})$ in a finite-dimensional
space $\cal{H}$.
Let $s_j = \{ P^{(i)}_j : i = 1,2,\ldots , n_j \}$, and let
$t = \{ Q^i : i = 1,2, \ldots , n \}$.
Then for any $i_1 , \ldots i_l $ and $i$ in the given ranges, we have
\eqn\eigenmatrix{
Q^i P_l^{i_l} \ldots P_1^{i_1} \rho^{1/2} =
\cases{
P_l^{i_l} \ldots P_1^{i_1} \rho^{1/2} \, , & {\rm ~if~}
$p(i_1 , \ldots , i_l , i) \neq 0 \, ;$ \cr
0 \, , & {\rm ~if~} $p(i_1 , \ldots , i_l , i) = 0 \, .$ \cr}}
\medskip\noindent
{\it{Proof}} \qquad The case when $p(i_1 , \ldots , i_l , i ) = 0$
is immediate.  But by triviality, for each $\{ i_1 , \ldots , i_l \}$
there is only one $i$ with $p(i_1 , \ldots , i_l , i ) \ne 0$, and  
since $Q^i = 1 - \sum_{j \ne i} Q^j$ the result follows. 
\medskip\noindent
\medskip\noindent
This implies that, if a projective decomposition occurs more than once
in a consistent set, there is at least one class of
consistent extensions which {\it can} be made to all consistent
extensions:
\medskip\noindent
\medskip\noindent
{\it{Lemma 7}} \qquad
Let $\S = ( \rho , \{ s_1 , \ldots , s_j , t , t_1 , \ldots , t_l , t,
s_{j+1}, \ldots , s_k \} ) \equiv (\rho, \{ S_1 , t, T , t, S_2 \} )$ be a
consistent set in which the projective decomposition $t$ is repeated.
Let $\S' = (\rho , \{ S_1 , t , T_1 , t , T_2 , t , S_2 \} )$ be an
extension of $\S$ by a further repetition of $t$ at some point between the
first two, so that $\{ T \} = \{ T_1 , T_2 \}$.
Then $\S'$ is also consistent.
\medskip\noindent
\medskip\noindent
{\it{Proof}} \qquad
Let $ \alpha , \beta , \gamma $ denote histories from $S_1 , T , S_2$
respectively, and $a,b$ projections from $t$.
Since $\sum_{\beta} p( \alpha , a , \beta , b , \gamma ) =
p( \alpha , a , b , \gamma ) = \delta_{ab} p( \alpha, a, \gamma )$, we
have that $p( \alpha , a , \beta , b , \gamma ) =
\delta_{ab} p( \alpha , a, \beta, \gamma )$, so that
the consistency of $\S$ implies that it is a
trivial extension of $(\rho, \{ S_1 , t, T , S_2 \} )$.
Likewise the set
$(\rho, \{ \bar{S}_1 , t, \bar{T}, t, \bar{S}_2 \} )$ is a trivial
consistent extension of $(\rho, \{ \bar{S}_1 ,  \bar{T}, t, \bar{S}_2 \})$,
for any subsets $\bar{S}_1 , \bar{S}_2 , \bar{T}$ of $S_1 , S_2$ and $T$.
In particular, $(\rho, \{ S_1 , t, T_1  ,t \} )$ is a trivial consistent
extension of $(\rho, \{ S_1 , t, T \} )$ for any subset $T_1$
of $T$.
Let $P^{1}_{\alpha}$ and $Q^1_{\beta}$ denote general
histories from $S_1$ and $T_1$ respectively, and
let $P_a$ denote a general projection from $t$.
Lemma 6 implies that
\eqn\xxx{
P_a Q^1_{\beta} P_b P^{1}_{\alpha} \rho^{1/2} =
 \delta_{ab} \, Q^1_{\beta} P_b P^{1}_{\alpha} \rho^{1/2} \, . }
Hence the consistency conditions for $\S'$ follow from those for $\S$,
as required.  Note that, by the same argument, any consistent
fine graining of $\S$ can be consistently fine grained by a further
repetition of $t$ between the first two.
\medskip\noindent
\medskip\noindent

We conclude with a short list of counterexamples.
By Lemma 7, a consistent
set in which a projective decomposition $t$ is repeated can
always be extended consistently by
adding an arbitrary number of repetitions of
$t$ between the originals.
However, Lemma 5 shows that adding $t$ after all existing
occurrences of $t$ is
not possible in all consistent extensions.  We cannot even, in general,
add $t$ to the past of all the original $t$'s:
\medskip\noindent
\medskip\noindent

{\it{Example 4}}  \qquad As in example 3, we take a three-dimensional Hilbert
space, with pure initial state $\rho$, and consider the one-dimensional
projections $P(z_1 , z_2 )$ defined by \pztwo.
Defining histories by the series of projective decompositions
$\sigma_i = \{ P^{(1)}_i , P^{(2)}_i \}$ for $i = 1,2,3$, where
\eqn\csfour{\eqalign {P^{(1)}_1 = P(0 , x_{12} ) \, , \qquad  &
P_1^{(2)} = I - P^{(1)}_1 \, , \hfill \cr
P^{(1)}_2 = P(0 , 0 ) \, , \qquad &
P_2^{(2)} = I - P^{(1)}_2 \, , \hfill \cr
P^{(1)}_3 = P(0 , x_{12} ) \, , \qquad  &
P_3^{(2)} = I - P^{(1)}_3 \, . \hfill \cr }
}
Again we see from \fcsthreeconds\ that, while the set
$( \rho, \{ \sigma_2 , \sigma_3 \} )$ is consistent, its
extension by past repetition, $( \rho, \{ \sigma_1 , \sigma_2 , \sigma_3 \} )$,
is not consistent if $x_{12} \neq 0$.
\medskip\noindent
\medskip\noindent
Another interesting question is the extent to which the usual notion
of correlated subsystems carries over into the consistent histories
formalism.  The answer is that, while one can find a consistent
set which describes the correlation, one can also find consistent
sets which are incompatible with the correlation, in the sense that
they have no consistent extension in which the correlation can be
displayed:\medskip\noindent
\medskip\noindent
{\it{Example 5}} \qquad Consider the Hilbert
space ${\cal{H}} = {\cal{H}}_1 \otimes {\cal{H}}_2$,
where ${\cal{H}}_1$ and ${\cal{H}}_2$ are
two-dimensional spaces with orthonormal bases $\{ v_1 , v_2 \}$ and
$\{ w_1 , w_2 \}$ respectively.  Take
\eqn\exsix{\rho =  p \ket{v_1 } \ket{w_1}  \bra{v_1 } \bra{w_1} +
          (1- p)  \ket{v_2 } \ket{w_2}  \bra{v_2 } \bra{w_2} \, ,}
define the projections $P_i = \ket{v_i } \bra{v_i }$,
$Q_i = \ket{w_i } \bra{w_i }$, $i = 1,2$ and
$Q_3 = {1 \over 2} (\ket{w_1} + \ket{w_2} )(\bra{w_1} + \bra{w_2} )$.
Let
\eqn\decomps{\eqalign{
\sigma_1 & = \{ P_1 \otimes Q_1 ,  I \otimes I  - P_1 \otimes Q_1 \} \, ,\cr
\sigma_2 & = \{ P_1 \otimes I , I \otimes I  - P_1 \otimes I \} \, , \cr
\sigma_3 & = \{ P_1 \otimes Q_3 , I \otimes I  - P_1 \otimes Q_3 \}
\, .}}
Then the set
\eqn\setone{
 \S_1 = ( \rho , \{ \sigma_1 , \sigma_1 , \ldots , \sigma_1 \} ) }
is consistent and describes the correlation of the subsystems corresponding
to ${\cal{H}}_1$ and ${\cal{H}}_2$, in the sense that the reduced 
density matrices corresponding to the histories of non-zero
probability are 
\eqn\reddm{\eqalign{
p^{-1} R_1 \ldots R_1 \rho R_1 \ldots R_1  &= \ket{v_1 } \ket{w_1}  
\bra{v_1 } \bra{w_1} \, ,
\cr
(1-p)^{-1} 
R_2 \ldots R_2 \rho R_2 \ldots R_2 &= \ket{v_2 } \ket{w_2}  \bra{v_2 }
\bra{w_2} \, ,}}
where $R_1 = P_1 \otimes Q_1$ and $R_2 = (I \otimes I  - P_1 \otimes Q_1)$.
However, the set
\eqn\settwo{
\S_2 = ( \rho , \{ \sigma_3 , \sigma_2 , \sigma_2 , \ldots , \sigma_2 \} ) }
is also consistent.  $\S_2$ does not describe the correlations, and
cannot be consistently extended to a set which does:
any set of the form
\eqn\setthree{
\S_3 = ( \rho , \{ \sigma_3 , \sigma_2 , \ldots , \sigma_2 ,
\sigma_1 , \sigma_2 , \ldots , \sigma_2 \} ) }
is inconsistent.\medskip\noindent
\medskip\noindent
This simple result suggests, in particular,
that one cannot generally deduce the quasiclassical
behaviour of one subsystem from a quasiclassical description of
a disjoint subsystem, since if the descriptions are for a sufficiently
short time then the effect of the interaction hamiltonian can
be neglected.
Put picturesquely, it is possible, given suitable initial
conditions, and restricting oneself to a sufficiently short time
interval, to find consistent sets in which the
earth is described quasiclassically, and
the moon is not, and in which no consistent fine graining allows one
to recover a quasiclassical description of the moon.
We shall suggest later that a similar problem arises for long time
intervals.

Suppose we accept the probability rules \probtwo\ for consistent histories.
We can still ask whether there might perhaps be some other, presently
unknown rule which assigns probabilities to general inconsistent histories,
in such a way that the probability sum rules hold.
As Goldstein and Page\refs{\goldsteinpage} have stressed,  
standard no-local-hidden-variables theorems\refs{\bellthm}
show that this cannot be done.  For example: 
\medskip\noindent
\medskip\noindent
{\it{Lemma 8}} \qquad  There exist inconsistent sets of projective
decompositions with the property that there is no
probability distribution on their histories from which one can derive
the standard probabilities \probtwo\ for all histories belonging
to consistent subsets.
\medskip\noindent
\medskip\noindent
{\it{Proof}} \qquad With the Hilbert space and density matrix of Example 2,
and using the notation defined there, consider the inconsistent
set $\S = ( \rho, \{ \sigma_1 , \sigma_2 , \sigma_3 \} )$,
where $\sigma_i =  \{ P_i^{(1)}, P_i^{(2)} \}$, with
$P_i^{(1)} = P ( i )$ and $ P_i^{(2)} = P ( - 1 / i )$ for $i=1,2,3$.
Let $a,b,c$ --- each taking the values $1$ or $2$ --- label the
projections in the various decompositions.
Each length two subset of $\S$ is consistent, so that any assignment of
probability weights $p(a,b,c)$ to the histories $P_1^{(a)} P_2^{(b)} P_3^{(c)}$
of $\S$ would have to satisfy
\eqn\lemmaviii{
p(1,1,1) + p(1,1,2) = \Tr ( P_2^{(1)} P_1^{(1)} \rho P_1^{(1)} )}
and eleven similar equations.  From the general
solution, which involves one free parameter,
we find in particular that
$p(1,2,1) + p(2,1,2) = - 1/25$, so that
at least one of the hypothetical probability weights must be negative.
\medskip\noindent
\medskip\noindent
As we shall explain
in the next two sections, the various results we have listed
are useful in clarifying the basic features of the consistent histories
formalism and its
interpretations.
They also illustrate that very little is known of the
properties of consistent sets even in finite-dimensional spaces,
and there are many simple unanswered questions.
How far, for instance, can the bound of Lemma 1 be improved?  What is the
length of a typical consistent set?  Or the expected length of
a consistent set constructed by choosing the first projective decomposition,
and each successive extension, randomly (using some natural measure on
the generalised grassmannian spaces) from amongst the consistent alternatives?
Can Lemmas 2--5 be extended to arbitrary initial density matrices?
Does a consistent set
$\S = (\rho, \{\sigma_1,\dots,\sigma_l \})$,
which is not maximally extended, always have an extension in the future,
$\S' = (\rho, \{\sigma_1,\dots,\sigma_l , \sigma_{l+1}, \ldots , \sigma_L \})$,
which is maximally extended,
or is it possible to find examples where non-trivial extension requires
projective decompositions to be inserted within the existing sequence?

We expect that in practical applications, where the space of
possible extensions tends to be of much
larger dimension than the number of consistency equations, then the naive
counting arguments are qualitatively correct.
That is, we expect that under these conditions
the solution space of non-trivial consistent extensions generically
contains manifolds
of large dimension.  Likewise, in practical applications the sets under
consideration tend to be very far from fully fine grained, and we expect
the conclusion of Lemma 2 to hold generically, whether or not the initial
density matrix is pure.

In this section we have applied the condition that the
objects of interest in the theory are the fundamental consistent
sets. From now on we shall relax the equivalence relations and the
condition of full fine graining.
The description of
IGUSes and quasiclassical domains which we shall need are
most straightforwardly made
assuming
both that the projection operators are assigned
times (contrary to our adopted equivalence relation) and that we can make
trivial consistent fine grainings.  The latter are needed since one of the most
important sorts of prediction we want to make
is of classical deterministic behaviour which is most familiarly
described in terms of repeated trivial extensions.

\newsec{Quasiclassicality and Approximate Consistency}

It is not obvious from our discussion so far that the consistent histories
formalism is able to give any useful description of the world we see, or
the laboratory experiments we do.  The case that it can relies on a study
of what Gell-Mann and Hartle have called the {\it quasiclassical} projection
operators describing our own classical domain: operators such as the integral
of the mass density, or the density of chemical species, over small regions,
the approximate position of an apparatus pointer, or the approximate current
through a photomultiplier.
As many authors have 
% refs on one line seems to make TeX happier
stressed,\refs{\bohmbook,\ \jooszeh,\ \zurek,\ \caldlegg,\ \gmhprd}
interactions with the environment are an impressively efficient mechanism
for quantum decoherence, even when that environment consists only of the
cosmic microwave background radiation.
Hence quasiclassical projection operators obey the consistency
equations \decohgmh\ to an extremely good approximation.
However, since no natural decoherence mechanism is perfectly efficient,
none of the quasiclassical projection operators listed above are ever likely
to precisely obey the consistency equations; nor is it easy (even in simple
models) to find quasiclassical projections which do.
In the context of the consistent histories formalism,
this point was first raised by Griffiths,\refs{\griff} and has been discussed
in more detail by Gell-Mann and
\hbox{Hartle.\refs{\gmhsantafe,\ \hartleone}}  Gell-Mann and Hartle suggest
that, in applying the formalism, approximately consistent sets should be
accepted for the purpose of making predictions, so long as the failure of
consistency (and hence of the probability sum rules) is so small as to
be experimentally undetectable.

This proposal has excited lively controversy.
Clearly, the use of approximately consistent sets does
raise new problems.  Most obviously, the problem of set selection is worsened,
since the class of approximately consistent
sets is larger than the class of consistent sets.
More fundamentally, many physicists would prefer a
theory in which probabilities are precisely defined and precisely obey the sum
rules.  The consistent histories formalism does, after all, have a natural
mathematical structure: why weaken it with ad hoc
prescriptions?\foot{Would similar violations of
Lorentz invariance be acceptable?}
Moreover, the formalism has at least one natural
interpretation in which it is impossible to ascribe a fundamental role to
the approximately consistent sets.\foot{This is the many-histories
interpretation described in section 5.}
Finally, it is often thought a sign of
weakness if one needs apparently contingent facts to make sense of a
fundamental theory: the overwhelming efficiency of environmental decoherence
and the impossibility of performing an infinite sequence of identical
experiments are both contingent on the hamiltonian and boundary conditions,
whereas the consistent histories formalism itself applies to any
hamiltonian and boundary conditions.

There are counterarguments for each of these points, and
Gell-Mann and Hartle certainly do not regard any of these problems as
serious weaknesses.  Their defence rests on a particular
view of the scientific enterprise and the role of
probabilistic theories: expositions can be found in Refs.
\refs{\gmhsantafe,\ \hartleone}.
We do not, however, want to enter this debate:
rather, we suggest that it is most likely irrelevant.  A typical analysis of
decohering quasiclassical projection operators applies a relatively short
series of relatively coarse grained projections to a system whose
Hilbert space is large or infinite-dimensional.  For example, an analysis
of a photon two-slit experiment might use a few fairly coarse grained
projective decompositions of the local densities of chemicals in a
photographic plate, at times either side of the arrival of the photon at the
plate, while the relevant Hilbert space would describe
a mesoscopic collection of particles.
Unless an implausible number of partial
derivatives vanish, this suggests that if the analysed set $\S$ satisfies
equations \decohgmh\ approximately then there is a large-dimensional
manifold of exactly consistent sets including points near to $\S$,
since the generalised grassmannian spaces characterising sets of
projective decompositions with ranks those of $\S$ will be
of dimension much larger than the number of consistency equations.

More precisely, for sets of the same length whose projection operators have
fixed ranks, one can define a metric by, say, taking
\eqn\metric{
d( \S , \S' ) = \sum_{j=1}^l \sum_{i=1}^{n_j}
d( P^{(j)}_i  , {P'}^{(j)}_i ) \, ,}
where $d(P, P' )$ is the operator norm of $(P- P' )$.
Then if $\S$ is consistent to within a small parameter $\epsilon$, one expects
to find an exactly consistent set $\S'$ (of the same length and with the same
collection of ranks) within a distance $A \epsilon$ of $\S$, where $A$ depends
on partial derivatives of the decoherence functionals at $\S$ and can in
principle be estimated.

It is crucial for this argument that the analysed set $\S$ is far from
being fully fine grained.  
Given a pure initial
density matrix in a space of dimension $n$,
one can construct examples in
which an approximately consistent set with no approximately 
consistent non-trivial fine graining contains more than $n$
non-trivial histories, and so cannot possibly be approximated by an
exactly consistent set.\foot{We thank Matthew Donald, to whom this
point is due.}
Thus, roughly, the conjecture is that for any
finite-dimensional Hilbert space $\cal{H}$
there is some positive bound $\epsilon (M)$, defined for finite
integer $M$, such that a generic set in the space of sets of 
histories, having
$M$ histories and approximately consistent to
order $\epsilon < \epsilon (M)$,
contains an exactly consistent set in its neighbourhood; 
that $\epsilon(M)$ is large enough that the approximately consistent 
sets of physical interest are well approximated by exactly consistent
sets; and that this last fact holds true for histories defined
on infinite-dimensional separable Hilbert spaces.

More precise mathematical statements (which neighbourhood?  
what bounds can be found for $A$?) should probably be  
formulated.  (There is some progress in this
direction.\refs{\mcelwaine})
It would also be 
very interesting to see some detailed investigations in models.
Still, for the moment we see no strong case for arguing 
that approximately consistent sets need be given
any fundamental role, and we
consider only exactly consistent sets hereafter.

It is worth mentioning here another of Gell-Mann and Hartle's
suggestions: the possibility that a single quasiclassical domain,
or a small number of such domains, might emerge from quantum
cosmology.
The point here is that at present the notion of a quasiclassical
domain is only an intuitive one:\medskip\noindent
\medskip\noindent
``Roughly speaking, a quasiclassical domain should be a set of
alternative decohering histories, maximally refined consistent with
decoherence, with its individual histories exhibiting as much as
possible patterns of classical correlation in time.  Such histories
cannot be {\it exactly} correlated in time according to classical
laws because sometimes their classical evolution is disturbed by
quantum events.  There are no classical domains, only quasiclassical
ones.'' \refs{\gmhquasi} \medskip\noindent
\medskip\indent
Gell-Mann and Hartle here consider maximally refined --
and not fully fine grained -- sets since
a quasiclassical domain is supposed to involve many trivial extensions,
so that the redundancy needed to characterise classical variables is
built into its description.

It would be preferable to have a precise notion of
quasiclassicality --- a calculable quantity,
defined by some sensible, but as yet unknown, measure.
The hope is that {\it the} quasiclassical domain or domains would be
given by one or more maximally refined
consistent sets with a
much higher degree of quasiclassicality than all the
others:\medskip\noindent
\medskip\noindent
``We wish to make the question of the existence of one or more
quasiclassical domains into a {\it calculable} question in quantum
cosmology
and for this we need criteria to measure how close a set of histories
comes to constituting a `classical domain'.'' \refs{\gmhquasi}\medskip\noindent
\medskip\noindent
``It would be a striking and deeply important fact of the universe if,
among its maximal sets of decohering histories, there were one roughly
equivalent group with much higher classicities than all the others.
That would then be {\it the} quasiclassical domain, completely
independent of any subjective criterion, and realized within
quantum mechanics by utilizing only the initial condition of the
universe and the hamiltonian of the elementary
particles.'' \refs{\gmhigus}\medskip\noindent
\medskip\indent
This last proposal can be fleshed out a little, under the assumption
--- which finite-dimensional calculations support --- that the fully
fine grained (and hence the maximally refined) consistent sets
are parametrised by algebraic curves.  One would
expect any measure of
quasiclassicality to be continuous, so that the most that could be hoped for
is that there is a consistent set $\S$ on
which the measure attains a global maximum, with the consistent sets in its
neighbourhoods arbitrarily close to this maximum quasiclassicality, and
which is much higher than any other local maxima.
Gell-Mann and Hartle's suggestion is then that all
those sets in the neighbourhood of $\S$ whose quasiclassicality is near
that of the maximum should
describe very similar domains, so that
to make predictions significantly different from those of
$\S$ one would have to use sets of much lower quasiclassicality.
The notions of ``similar'' and ``near'' need to be made
precise here.  Very roughly speaking, the hope seems to be that
the relevant matrix of second derivatives is positive definite and
that its eigenvalues have some lower bound.  Finite-dimensional
calculations give no clue as to the plausibility of this proposal, but
their complexity does suggest that, even if the relevant quantities
can be defined, it may be rather hard to test.

\newsec{Interpretations of the Formalism}

Having set up the formalism and shown that there will be very many consistent
sets of histories in even the simplest models, we must now consider how to
make contact between the theory and our observations.
In the consistent histories formalism, 
there is no distinction between the
consistent sets in a given model of the universe. True,
most of the sets will describe the history of the universe in terms of
very complicated and non-local projection operators and it may happen that
one or more of the sets contains histories which are described by
projections that are much more familiar to us, but the formalism ascribes
the same validity to them all.  The problem appears only to be
worsened when the formalism is applied to relativistic 
quantum field theory. 
How, then, are we to make sense of the
formalism?  Or, to put things more practically, if we are given a quantum
cosmological theory, how (and to what extent) can we check whether it
is right?

\subsec{Cosmological Tests}

There are two essentially distinct tests of a quantum cosmological theory:
does it describe the past? and can it predict the future?
Neither question is as simple as it seems.
We first consider the problem of accounting for the past.

It seems obvious that any good theory must be consistent with what
Omn\`es calls the {\it actual facts} --- the events we regard as 
having actually been observed (in short, the data).  
That is, the projection operators corresponding to our
actual experiences must comprise a history belonging to at least one
consistent set.  Let us call this the {\it actual history}.
Further, the conditional probabilities
assigned by the formalism must be borne out by what we have seen.
For example, our two-slit experiments should have produced roughly the right
interference patterns.

The difficulty here is that the notion of actual
history is ambiguous.\foot{We thank Andy Albrecht for discussions of
this point.}
This raises unavoidable questions about the
proper scope of a cosmological theory.
One could, solipsistically but quite consistently,
include only one's own experiences in the actual history, or one could try to
include those reported by others.  One could try to describe only the present
state of one's mind, or one could assume that one's memories are essentially
accurate and try to construct a history corresponding to the past events which
one recollects.  More generally, one can take what appear to
be historical records --- Darwin's notebooks, the fossil record,
the microwave background --- merely as present facts to be described by
present-day projection operators, or as prima facie evidence for past facts
which should be included in the actual history.
These are live issues in quantum cosmology.  Of course, no one
seriously supposes that we shall be faced with
rival cosmological theories, one of which claims that Darwin's notebooks were
written in the ordinary way, and the other that our present day records
of Darwin's observations originate from a fortuitous quantum fluctuation
in the neighbourhood of the Cocos Islands on 1st April 1836.
But we cannot interpret the formalism sensibly, or consider how it might
best be developed, without taking a definite view on these points.
Indeed, the entire formalism
is unnecessary if one is content only to
account for the ``marvellous moment'' and to treat all memories and
records simply
as statements about the present.\foot{Bell seems to have been the first
to discuss this proposal in the
context of many-worlds interpretations.\refs{\bell}
As he mentions, a well-known variant on the ``marvellous moment'' philosophy
accepts records of recent history but forbids historical reasoning
beyond a certain point in the past.\refs{\gosse}}
This ``solipsism of the moment,'' which is
argued against by Hartle,\refs{\hartlemarvellous}
is the most extreme anti-historical position.
If one takes this line, everything can be
described by a single projective decomposition now.
Theories of the initial density matrix or the Hamiltonian are then
simply data compression algorithms: one does physics (or rather, since
processes extended in time have no meaning, has the illusion of doing
physics) in an attempt to describe more succinctly one's momentary
experience.

Our own view is that quantum cosmological theories ought to allow
ordinary quasiclassical historical reasoning during the era in
which, according to conventional cosmology, the universe was quasiclassical.
Hartle makes the point well:\refs{\hartlepast}\medskip\noindent
\medskip\noindent
``The simple explanation for the observed correlation that similar dinosaur
bones are located in similar strata is that dinosaurs {\it did} roam the earth
many millions of years ago and their bones are persistent records of this
epoch.  It is by calculating probabilities between the past and different
records today that correlations are predicted between these records.  It is
by such calculations that the probability for error in present records is
estimated.  Such calculations cannot be carried out solely on one spacelike
surface.''\medskip\noindent
\medskip\indent
Of course, memory can deceive and records can mislead, and for these
reasons alone there may never be complete agreement on
what the actual facts are.
Moreover, as we shall discuss presently, there are real theoretical
dilemmas in selecting the actual facts.
There are also differences of opinion over how, or perhaps even
whether, criteria of beauty and simplicity should be applied to select
cosmological theories.

Let us put these difficulties aside for the moment,
assuming that the actual facts are given and restricting
attention to theories which are agreed to be suitably simple and
beautiful.
We believe that a successful cosmological
theory ought to describe the vast bulk of our own and others'
experiences, and the past events reconstructable from records
available to us, as events in a history from a
single consistent set.
Although we hope that most would concur with this, we are not
sure there is a consensus.
The formalism certainly allows a wide range of possible views.
At one extreme, we have pure solipsism of the moment: advocates of
this view will accept any theory that describes their present
experience.
Next we have recent history solipsism: this accepts any theory which
describes actual facts over some period in time up to and including
the present.  Some recent history solipsists might simply want to describe
their own experiences, and see no need to include events occurring
before their birth; others might choose some other interval.
A third position is that any theory must allow us to describe
historical events at arbitrarily early times, but not necessarily in
the same consistent set.  It is hard to produce convincing examples
but, suspending disbelief for the sake of illustration, one might imagine a
quantum cosmological theory in which it
was possible to describe early time matter density anisotropies leading
to galaxy formation but
{\it not} possible to describe all the relevant anistropies in the same
consistent set.
Finally, one can demand that the theory
describe, within one history of a single consistent set, essentially
all the actual facts implied by quasiclassical historical reasoning.

We ourselves subscribe to this last position.
The views of the developers of the consistent histories approach will
be examined in later subsections.
Meanwhile we should note that the extracts we have cited from Hartle's
papers have to be interpreted with great care.
Read naively, as statements in ordinary language, they set out a still
stronger position.  They demand that a cosmological theory should give
a clear account of the past which unambiguously states that, with
probability very close to one, key historical events actually took place.
As it happens, this is indeed our view, but it is not Hartle's:
as we shall explain, the cited extracts are {\it not} meant in
their ordinary sense.

Now, if we adopt any position other than solipsism of the moment, we must
explain what criteria we would use to identify the actual facts.
It is a complicated question.
The assumption of quasiclassical behaviour some way into the past seems to be
part of the answer, but only part.
It is closer to the truth to say that we would first build and
test incomplete, high-level cosmological
theories --- describing the evolution of
terrestrial life, the formation of the solar system, or the seeding of
anisotropies from which galaxies develop --- and then use successful theories
of this type as actual facts to be explained by a fundamental quantum
cosmology.\foot{
Although we think it unlikely that there could be serious conflict
between the two levels of theory, one could perhaps imagine some
tension at the boundaries of quasiclassicality.
Thus, while it seems inconceivable that there could be
any quantum cosmological theory, consistent with the present data, in
which no consistent set contains a quasiclassical description of
terrestrial evolution, it does seem conceivable (if perhaps only as a
result of our present ignorance) that the desire to
maintain some elegant cosmological theory of structure formation might
conflict with the desire to adopt the simplest possible theory
of the boundary conditions.  It is hard to see how any
general algorithm could be found to decide which theory should be
rejected.}
This procedure is very interesting.
It would be good to see a serious examination of exactly what is involved
and how it can be justified since, if quantum cosmology ever produces testable
theories, we shall need to understand precisely which tests we are
applying, and why.

Discussions of actual facts, then, must be understood in the context
of some high-level cosmological theory or theories.  We now take
this for granted, and refer to {\it the} actual facts, and we shall
assume these are to be included in one history of a single consistent set.

The need to account for the actual facts
in the consistent histories formalism
gives a strong constraint on quantum cosmologies.
It may well be impossible to satisfy this constraint in some, or all,
versions of the formalism.
For example, even assuming that the basic ideas of the consistent histories
formalism and of quantum cosmology are correct, one can well  
imagine that the actual facts
cannot be described by a history within a consistent set in the
projection operator formalism that we consider in this
paper, and that, as Gell-Mann and Hartle
suggest,\refs{\gmhonbranches} we really need to consider
histories belonging to a branch-dependent consistent set ---
that is, one in which the projective decompositions after any time $t$ are
allowed to depend on the history selected up to that 
time.\foot{Why, for instance, should a microscopic difference in 
local densities of 
chemical species in a laboratory experiment necessarily decohere in 
histories in which the solar system was never formed?} 
In other words, the various proposals in the consistent histories
program already constitute a collection of theories of nature, and these
theories are in principle falsifiable.
However, since all of these alternative proposals only exacerbate
the problems we discuss in this section, we shall continue to restrict
our attention to consistent sets of histories defined by
branch-independent projection operators.
Note also that, as the proposals
considered in section 7 illustrate, it remains
possible that the fundamental principles of quantum cosmology (the
description of
the universe by hamiltonian evolution from quantum initial and final
conditions) are correct, but that no consistent histories
formalism is.

We turn now to the future.  Here we must be careful with our language, since
we shall see later that the formalism has different interpretations, in which
the discussion has to be phrased differently.  The point, though,
is that we can make predictions about the future only once a particular
consistent set, incorporating the actual history, is specified.
In particular, if we want to predict --- as we surely do --- that we shall
continue to experience a quasiclassical world, we need a consistent set
in which this is described as following, with a high conditional
probability, from the actual history.
How good --- that is, how strongly predictive --- the cosmological theory is
very much depends on whether we can find an argument which genuinely
depends only on the consistent histories formalism, the density matrix, and
the particular interpretation of the formalism which we are using, and which
selects out the relevant consistent set and makes clear why it is the
one we must use.
This, we suggest, is the crucial point on which, in the end,
interpretations of the formalism stand or fall: we continue the discussion in
the subsections below which deal with the various possible interpretations.

We have still to explain what we mean by an interpretation
of the formalism, and why it requires one.
By an interpretation we simply mean a clear set of rules which explain how,
in a general cosmological model, we are supposed to translate mathematical
statements about the formalism into statements about physics.
In this sense, of course, any mathematical theory of physics needs an
interpretation.  We suspect that the reason why physicists are often
loath to discuss problems of interpretation is that, for many interesting
theories, it seems that nothing is to be gained by discussion.  There
seems to be
general agreement on how to make sense of classical mechanics, or
general relativity, so that there appears to be little need to set out
precisely what it is that we agree on.
No such agreement yet exists for the consistent
histories formalism.  There seems to be real confusion about what the
formalism means or is intended by its authors to mean.
Many arguments in the literature rely on some
notion of interpretation, and it is far from clear that all these arguments
can actually simultaneously be made in any single interpretation.

In the remaining subsections we shall try, as far
as possible, to find interpretations in which the key arguments in the
literature can be justified.
We shall see that there are
at least four distinct interpretations of the formalism.
Each interpretation is a serious attempt to make sense of the
formalism; unfortunately, each has problems.

\subsec{Omn\`es }

Omn\`es  uses the formalism as a
way of analysing the logical consistency of various
collections of statements, reaching an interpretation of
the formalism which is logically impregnable.
As Omn\`es points out, this interpretation
allows an analysis of all the propositions which the Copenhagen 
interpretation would produce, and a great many more. 
However, as we shall explain, this involves a loss of predictive power.
Omn\`es' approach is deliberately cautious but
uncomplacent in spirit.  Omn\`es attempts to describe the minimum of logical
structure necessary to make sense of the formalism but, as
he has stressed,\refs{\omnespriv} his approach
is not intended to exclude other interpretations,
nor is it meant to discourage proposals to extend the formalism.
Omn\`es' own primary concern is with quantum mechanics rather than quantum
cosmology.  We believe, however, that he has given a sensible procedure for
evaluating quantum cosmological theories.
We shall try to sketch the method below:
for a less superficial treatment we recommend Omn\`es' thoughtful and
detailed exposition.\refs{\omnes}

Suppose we have some cosmological theory which passes the first test for
viability: that the actual facts are described by a history from
at least one consistent set.
We will certainly want to use the theory to make predictions.
We may also want to make statements about the past.  It is useful
to divide these into two types: statements about events before
any of the actual facts, and statements about events occurring at times
in between actual facts.
Past statements cannot be used to test the theory, but they are of interest
nonetheless, and we shall use the terms {\it retrodictions} and
{\it paradictions} to refer to the two cases.\foot{The term retrodiction
has generally been used to apply to both, but it seems useful to make
the distinction even in the non-relativistic formalism.
In the relativistic case, some such distinctions seem essential.
If the actual facts describe projections
localised within a particular region $R$ of spacetime, we clearly want
to refer respectively to derived statements describing propositions in
regions which
are in the future and past light cones of all points in $R$ as
predictions and retrodictions.
We refer to all other derived statements --- describing projections
which are not localised either entirely in the future or entirely in the past
of $R$ --- as paradictions.
It is probably then useful to refine the class of paradictions
further.  For example, defining the future boundary $\partial_f R$ of
$R$ to be the set of points in the closure of $R$ whose
future light cones do not intersect $R$, one can define
{\it weak predictions} to be statements referring to events lying in the
union of the future light cones of the points in $\partial_f R$, and
and similarly {\it weak retrodictions}.}
In principle, Omn\`es does indeed allow us to make predictions, retrodictions
and paradictions.  However, the only new statements which we
can strictly deduce are what he terms {\it true propositions} ---
those which hold
with probability one in
any fundamental consistent set fine graining the actual history.  We slightly
extend Omn\`es' discussion by calling these {\it definite} true
propositions and
further define a {\it probabilistic} true proposition to be a
proposition which holds
with some probability (not equal to zero or one) in every consistent 
fine graining.
There might possibly be surprising and non-trivial true propositions in
some theories.  However, in most theories the true propositions seem to be
very limited.  Let us consider the possibilities case by case.

The situation is clearest if the initial density matrix
is pure and if we consider future true propositions.
Let $\S$ be a consistent set including a history describing the
actual facts.  Suppose $P$ is a true proposition, definite or probabilistic.
Then $\{P, 1-P \}$ may be consistently
added to any consistent fine graining of $\S$. Lemma 5 shows that if this
is the case then $\S$ must be maximally extended. In other words, either
the future is entirely predictable, or there are no true future 
propositions.\foot{Note that this precise result assumes the 
consistency conditions \decohgmh\ .  It would be interesting to
see what difference other (weaker or stronger) sets of conditions
make.  The problems with Omn\`es' interpretation, however, are not
solved (nor claimed to be) by his choice of consistency conditions.}
This may seem counterintuitive.  One might like to believe that
future predictions divide up into classical and quantum events, with the
former inherently predictable and the latter intrinsically probabilistic;
and it is tempting to think that this means the classical events correspond
to definite true propositions, and the quantum events
to probabilistic true propositions.
Within a single consistent set this remains an attractive picture: trivial
consistent fine grainings give deterministic predictions and non-trivial
consistent fine grainings give
truly stochastic ones.
But, when all the sets are considered on the same footing, as they are
in the Omn\`es interpretation,
this is not the case (at least for pure initial states).
Either all future events are described by definite true propositions, or
none are
either definite or probabilistic true propositions.
It is logically possible that the unpredictability we ascribe to
quantum experiments is not genuine, and that in principle --- if only we knew
the full set of actual facts and the correct quantum cosmological theory ---
we could predict the outcome of all experiments with certainty.
However, unless one is willing to make this leap, unpropelled either
by evidence or argument, one is forced to accept that there are no future
true propositions.

We have no analogue of Lemmas 2 and 5 in the cases where the initial
density matrix is impure or where there is a non-trivial final density
matrix.  Presumably, future true propositions (either definite or
probabilistic) are not generally to be expected in
these cases either, unless (perhaps) the actual history's set is in some
sense close to fully fine grained.
In any case, given the current state of our understanding,
there seems no prospect of any future true proposition being identified.

Nor are past true propositions easy to come by.
Lemma 7 tells us that if the actual history
contains a repetition, so that the same projection $P$ describes events at
times $t_1$ and $t_2$, then further repetitions of $P$ at times $t$ with
$t_1< t < t_2$ {\it can} consistently
be included in all consistent fine grainings.
But these repetitive paradictions are the only general type of
past true propositions which we have been able to identify.
In particular, Example 4 shows that trivial past extensions, like trivial
future extensions, do not generally describe true propositions.  We know of
no examples of probabilistic past true propositions, either retrodictive or
paradictive.

Omn\`es was certainly
well aware that there are relatively few true propositions (beyond
those asserting the actual facts).
However, the two examples of true propositions identified in his
paper do not deserve that title.
Omn\`es uses the term {\it sensible logic} to describe any
consistent set which incorporates
the actual facts.  (Here {\it logic } simply means 
consistent set.) 
He argues that:\refs{\omnestrue}\medskip\noindent
\medskip\noindent
``Measurement theory $\ldots$
can be used to prove that the result of an experiment is always true.
Another example comes from
determinism: a past classical property that can be reconstructed logically
in a deterministic way from present records can be said to be true, even when
the sensible logics one is using involve only the
present facts.''\medskip\noindent
\medskip\noindent
and adds the comment:\medskip\noindent
\medskip\noindent
``It seems that the two examples just given are the only ones.  Thus one
recovers essentially Heisenberg's point of view as far as quantum events
are concerned, except for a deeper understanding of the meaning of truth
among the phenomena themselves.
The second example also answers an old question: It shows that a standing
object at which nobody is looking is still nevertheless at the same place,
and this can be taken to be true despite the fact that classical physics
relies upon quantum mechanics.''\medskip\noindent
\medskip\indent
This is very much the sort of result one would hope for.
Unfortunately, neither example is correct, as Omn\`es now
accepts.\foot{We are
very grateful to Roland Omn\`es for many helpful discussions on
this point.}
The first, in fact, needs no new discussion of ours.
Omn\`es' definition of an idealised measurement is an interaction between a
measured system $Q$ and a
macroscopic object $M$, taking place between an initial time $t_m$ and
a final time $t'_m$, with the usual correlative properties, so that when the
initial state of $Q$ is an eigenstate of some discrete observable $A$ with
eigenvalue $a_n$, the final state of $M$ is an
eigenstate of a pointer observable
$B$ with eigenvalue $b_n$.  The {\it result} of the experiment is ``a quantum
property, namely $A= a_n$, stating something about the measured system
(for instance, a value for a spin component).  It is a property holding at
the time $t_m$ when the interaction begins.''\refs{\omnesmeasure}
We can, for example, take $Q$ to be a spin-$1/2$ particle prepared in the
state $\sigma_x = 1/2$ at time $t_i < t_m$ and then left undisturbed until
the onset of the measurement interaction,
and $A$ to be the observable $\sigma_z$, and suppose
that the final state of $M$ points to the result $\sigma_z = 1/2$.
The actual history here is the final state of $M$ and the prepared state
of $Q$.  Now we can consistently extend the actual history to
include projections onto $\sigma_z$ at times $t$ such that
$t_i \leq t \leq t'_m$.  However, as
Griffiths points out,\refs{\grifftwo}
we can also consistently extend the actual history to include projections
onto $\sigma_x$ at time $t_m$, and we cannot consistently add a
projection onto $\sigma_z$ at time $t_m$ to this extension.
(Strictly speaking, this is trivially true, since the formalism does not
allow non-commuting projections to be applied at the same time.
A criticism more sympathetic to the spirit of the discussion is that
for ``$\sigma_z =1/2$ at time $t_m$'' to be regarded as true, it should at
least be possible to include the projection onto $\sigma_z$ consistently at
times $(t_m \pm \epsilon )$ for arbitrarily small $\epsilon$, and
this cannot be done for the negative choice of sign.)

Omn\`es' second example is particularly interesting.  In the first place,
one needs to establish some correspondence between classical determinism
and the triviality of the corresponding consistent sets.
Essentially, what is needed is to show that the
Heisenberg picture quasiprojection describing a classical system in a cell
of phase space at time $t_1$ is the same operator as the quasiprojection
describing the system in a deterministically evolved cell at some other
(earlier or later) time $t_2$.  Omn\`es gives
a careful discussion of the conditions under which this can be
shown.\refs{\omnesclass}
Let us assume that the correspondence has been established, and further
that we can replace quasiprojections by projections.
These are the most favourable possible assumptions, and they reduce
the statement to the claim that if a projection operator belongs to
the actual history, its past repetitions describe true propositions.
We have shown above that is not generally so; only when the actual facts
contain repetitions are there generally any true propositions, and in this case
we can only make further, repetitive paradictions.
In Omn\`es' language, one might --- with careful analysis --- be able to argue
that if a tree was first discovered standing in a forest yesterday, and is
still there today, and these discoveries are included in the actual
facts, then it is true that it continued to stand while
unobserved last night.
But it is not generally true that it did so last year, or, for
that matter, will do so
tomorrow.

Omn\`es' position on the role of historical reasoning in
identifying actual facts is unclear to us.
What is clear is that ``truth'' is not the 
criterion which we, as adherents to
ordinary historical reasoning, would use when building up the actual
facts.  For, once we have discovered
the tree, we may well wish to
include statements about its past classical history among the actual facts.
We may, for instance, want to deduce from its
growth rings that it has been standing for the past
thirty years, just as we want to deduce the past existence of live dinosaurs
from the fossil record.  These deductions are made using
classical determinism (or more accurately quasiclassical near-determinism)
rather than any criterion within the formalism.

In summary, an interpretation which relies
on the use of true propositions is unlikely to allow any
testable predictions.  While Omn\`es did not explicitly discuss
prediction, we believe he would agree this is a shortcoming in an
interpretation.
Our own view is that, because of the overwhelming variety of consistent
sets and the strength of the condition that a
proposition be includable in every
consistent extension of the actual history, it will not be possible
to resolve this particular problem without going beyond the formalism
of consistent histories.  
As we shall discuss later, it seems to us that the simplest
and most natural way to do this is to supplement the 
formalism by a set selection criterion.

However, remaining within the formalism of consistent histories,
one can still make testable predictions using Omn\`es' concept of
{\it reliable propositions} --- propositions which  have probability
one in at least one, but
not all, of the consistent sets which fine grain the set of the
actual history. Again we term
such propositions {\it definite} reliable propositions and
define {\it probabilistic} reliable propositions to be propositions which hold
with some probability not equal to one in at least one consistent
fine graining.
The statements physicists are accustomed to predicting are
reliable propositions.  For example, the predictions of classical
determinism are examples of definite reliable propositions and the
predictions of the
results of quantum experiments are probabilistic reliable propositions.
Unfortunately, as Lemma 2 makes clear, the same is
true of many more statements which we do not want to predict: if there are
probabilistic reliable propositions, then there are continuous
families of consistent
sets containing distinct reliable propositions.
If we want to predict that a Stern-Gerlach experiment will end up with
a pointer indicating the
result $\sigma_z = 1/2$ with probability $p \neq 0,1$, we have to accept
that this prediction belongs to a continuous family of predictions, each of
which is given equal status in Omn\`es' interpretation.  It is hard to say
precisely what all these predictions will be in a realistic laboratory setting,
but they would have to involve statements about macroscopic superpositions
of (at least) the pointer.

However, by considering the possible consistent fine grainings of the
actual history, we can make perfectly good {\it conditional} predictions
about the future.
Thus, {\it if}, for the
duration of an experiment,
a bubble chamber is described within the formalism by projection
operators delimiting the densities of chemical species within small
volumes, and if that description includes the appearance of a particle
track, then in the absence of external fields the track will be
approximately linear and
the probability distribution for
the track angle will be that given by standard quantum calculations.

Note, though, that it is almost certainly impossible to use the
formalism to make conditional predictions of the usual type, which involve
quasiclassical inferences.
For instance --- and here we go beyond the discussion after Example 5
--- we almost certainly cannot deduce that if the earth is in the
expected orbit throughout next Thursday then so will the moon be.
To make the deduction, one would have to show that, from the
actual facts describing the earth's trajectory quasiclassically,
the moon's orbit (as described by appropriate projection operators)
followed as a series of true propositions.
This seems theoretically implausible --- we have no proof, but doubt
that any non-trivial paradictions are true propositions --- and,
practically speaking, an impossible calculation.
To predict the moon's trajectory within the formalism, we need the
independent assumption that the moon will be described by the right
sort of projection operator.  If the projection operators are those
onto densities of chemical species within small volumes of the moon's
orbit, then they will describe the moon's expected orbit, whose
predictability follows from the quasiclassicality of those operators.
However, there are almost certainly other descriptions, consistent
with the quasiclassicality of the earth but inconsistent with that
of the moon.

Still, the conditional predictions which can be made are extremely useful.
We would no doubt be impressed if a cosmological theory were produced which
were consistent with the actual facts and which, in Omn\`es' interpretation,
made many correct conditional predictions, even though it
made no unconditional predictions.
This would nonetheless be a very weakly predictive sort of theory.
It would leave us puzzled as to why the world appears
quasiclassical and unable to predict that this
quasiclassicality will persist.
It remains a perfectly consistent and testable
scientific theory; it is logically possible that nature allows
no unconditional predictions and that, within its domain of validity,
an Omn\`es-interpreted cosmological theory is the best we can hope for.
But persistent quasiclassicality is such a basic feature of our
experience, and failure to predict it leaves such a significant gap in
our understanding of the world, that we feel that some argument, or
some new interpretation, or some new formalism, that would amend the situation
should be sought.

The only argument relevant to this point in the literature is that of
Gell-Mann and Hartle on the properties of IGUSes.  We shall come to it
later.  Note though, that it cannot be made in the language of actual
facts and true propositions.

\subsec{Griffiths}

In this subsection we very briefly sketch the views on interpretation that
Griffiths has set out in
a recent article.\refs{\grifflogic}  We here discuss Griffiths' proposals
only in so far as they relate to the problems of prediction with
which we are concerned; the reader is encouraged to
consult the original paper.\foot{
We shall not discuss retrodiction since, as far as we are aware,
Griffiths has set out no definite
position on the issue of historical reasoning and the nature
of the actual facts.}

To deal with the problem that different consistent sets allow different
descriptions of nature, Griffiths introduces
a set of rules for reasoning in quantum theory which he calls {\it Logic}.
Griffiths' Logic contains, in a precise way, all of Omn\`es' sensible logics.
However, Logic is a much weaker propositional calculus than ordinary
classical logic.  Most strikingly, the rules of Logic state that if
proposition $P$ is inferred with probability one in a sensible logic
and proposition $Q$ is inferred with probability one in a second,
incompatible, sensible logic then ``$P$'' and ``$Q$'' are separately
predicted but their conjunction ``$P$ and $Q$'' is declared meaningless.
Griffiths' case for Logic is, essentially, that it is a natural
way to interpret the consistent histories formalism; it may be a conceptual
weakening, but so (in some sense) were other advances in physics, such as
the abandonment of absolute time in special relativity.

It seems to us that this analogy does not hold.  Theoretical revolutions
typically involve radical conceptual revision rather than simple
weakening and, when they succeed, do so because of their greater
predictive power.  But on questions of prediction, Logic is of no help.
For, once again using Lemma 2, we see
that if there are non-trivial predictions
to be made at all and if the initial state is
such that the quasiclassical variables decohere,
then Logic predicts both that ``the universe will continue to be
quasiclassical'' and ``the universe will not continue to be
quasiclassical''.  The fact that this does not Logically imply ``the
universe will continue to be quasiclassical and the universe will not
continue to be quasiclassical'' saves the scheme from
self-contradiction but will not, in itself, be of 
comfort to many.\foot{Some may regard these statements as
alternative rather than contradictory descriptions.  Note though
that a Logical analysis of Lemma 8 or the standard no-local-hidden-variables 
results does produces direct contradictions.} 
Griffiths' view is that one needs not only the formalism (of which
Logic is a part) but also
a theory of human experience in order to make unconditional
predictions.\refs{\griffithspriv}
However, if one appeals to a theory of experience in order to
select a consistent set in which we can make unconditional
predictions, it is not clear that one has any scientific need for
a logical scheme relating the statements from distinct consistent sets.

In summary, the predictive content of Griffiths' interpretation is
meant to follow from an implicit theory of experience.  We shall consider
what this entails when we come to examine Gell-Mann and Hartle's ideas.

\subsec{Many Histories}

We now set out another interpretation of the formalism, which requires that
we take seriously the notion of many coexisting but non-interfering
histories.
This is inspired by Griffiths' interpretation, but
requires no change in the rules of logic: different histories peacefully
coexist.\foot{We are grateful to Bob Griffiths for mentioning this
interpretation in discussions.}
In addition to the changes needed to ensure this, we also introduce a slight
technical modification by insisting on sets which are exactly
(rather than approximately) consistent.

To set up this interpretation, we require that from each of the
fundamental consistent sets
$\S$ precisely one history $H(S)$ is chosen, the probability of any particular
history being chosen being precisely its probability $p(H)$, defined in the
usual way.
The interpretation then states that all of the chosen
histories, and no others, are realised.  The true description of nature, in
this interpretation, is the list of all the chosen histories
$\{ H(S) : S {\rm~a~fundamental~consistent~set} \}$, and each
history constitutes a complete description of one of an infinite 
collection of (for want of a better term) ``parallel worlds'' .  

Perhaps it is worth stressing a few details since, despite its simplicity,
this interpretation seems to cause confusion.  It is important that
just one history from each set is realised.  
If one postulates that 
all the histories of each set $\S$ are realised (as is sometimes
suggested \refs{\saunders}) then no role has been assigned to the
probabilities, and there seems no obvious way of introducing further
assumptions which would allow probabilistic statements to be deduced.
In other words, there is no precise analogy here with Everett's
suggestion \refs{\everett} that
all possible results of quantum measurements are always (in some sense)
equally realised:
this is a many-histories interpretation, but the many histories belong to
different consistent sets.\foot{
It should, nonetheless, be true that all possible
results of most quantum measurements are realised, but for a different
reason: most measurements can be described in
infinitely many different fundamental consistent sets
if Lemma 2 is any guide, or if our conjectures about the generic dimensions
of the solution spaces for the consistency equations are correct.
}

It is also crucial for this interpretation that exactly consistent sets
are used.  The statement that a history is realised is supposed here to be
an absolute statement of fact, quite independent of any anthropocentric
considerations about the accuracy to which we can test the history's
predictions.\foot{Attempts have been made to consider
approximate histories.\refs{\saunders}}

Some may feel uneasy at the use of a probability distribution
to select histories, or find it hard to make sense of
the idea that the chosen histories ``are realised''.
To take the second point first, we mean simply that the chosen histories
are supposed to correspond to nature; in just the same way, general relativity
admits a large class of four-dimensional manifolds, equipped with the
appropriate tensors, as solutions, of which one is supposed to correspond to
the universe we inhabit.   As for the use of probability, we see nothing
particularly distinctive about its use here: the relevant
debate is really not over the consistent
histories formalism at all, but over whether
any fundamental probabilistic theory of nature is meaningful or
acceptable.\foot{We are happy with such theories, but this debate is beyond
our scope.}

Still, the idea of describing reality by a collection
of sets of histories is unfamiliar and
perhaps uncomfortable.
We shall not argue that the picture is attractive, or particularly
plausible: we do claim, though, that the proposal makes sense.
In any sensible cosmological theory, the interpretation will produce a
highly structured set of
mathematical objects exhibiting complex patterns and it seems to us that,
although its character is rather different from familiar theories such
as general relativity, it therefore should be considered a
candidate mathematical theory of physics.  We now need to examine whether
our experiences, and the world they describe, could be reflected by
some subset of those mathematical objects, and if so,
whether that subset is identifiable a priori or only post hoc.

Here we are not optimistic.
The interpretation seems a very natural one, using as it does only the
basic notions of consistent set, history, and probability.
It also gives an unambiguous description of a definite reality, which
some like.  In practical terms, though, it is
indistinguishable from Omn\`es'.  Our own experiences are supposed to be
described by one of the realised histories from a given cosmological
theory.  To test this hypothesis, we
first have to be able to find some consistent history in which we can
describe our past and present experiences, and any historical
events which we decide we should deduce from records.  We have, in other
words, again to find Omn\`es' actual facts within the theory.
We then turn to predicting the future, and --- for the
reasons outlined above --- discover that we can make conditional
predictions, but cannot make unconditional ones.
Among the histories which are realised, there will be many which
describe the actual facts.  In most of these, quasiclassicality will not
persist in the future.  We are again left puzzled as to why
quasiclassicality does persist in the consistent history we actually
experience.\foot{This interpretation does perhaps give
truly hardened anthropicists a little more room for manoeuvre.
It might possibly be argued that some theory of
experience will be found which predicts that we have no awareness
in nonquasiclassical histories.  It could then be maintained
that an uncountable number of close (though imperfect)
facsimiles of ourselves do indeed perish in histories where
quasiclassicality dissolves, while we survivors are inevitably
left bemused by our fortune.
We do not ourselves take this type of argument seriously, but shall not pursue
the point --- which involves questions about the nature of identity and the
validity of anthropic reasoning that go beyond our scope.
We note, though, that any criticism of Gell-Mann and Hartle's position
--- which we shall discuss in the next subsection --- applies equally
to this argument, so that the extravagant multiplicity of facsimiles
and the invocation of anthropic principles seem to be drawbacks which bring
no compensation.}

\subsec{IGUSes and Quasiclassical Domains}

\leftline{\it 5.5.1 Preliminaries}
\vskip .1in
We now come to Gell-Mann and Hartle's seminal work on
the consistent histories formulation of quantum cosmology.
As in the earlier two subsections, our aim is to find a formal
interpretation that accords with Gell-Mann and Hartle's
views and in which we can draw their stated conclusions.
Here we must admit that we are in difficulties.
Gell-Mann and Hartle's writings on the consistent histories
formalism sparkle with ideas, proposals, suggestive observations 
and rhetorical arguments.  The totality of views expressed do 
not, it seems to us, constitute a coherent and internally 
consistent interpretation.
In part, this is a consequence of an evolving view: for example,
some of their most controversial interpretational
arguments are contained in an as yet unpublished paper, 
the current revised version of which\refs{\gmhcommnew} differs from the 
preprint version which is presently available on the 
bulletin boards.\refs{\gmhcomm} 
Here our criticisms refer in detail to the earlier 
version.\foot{
It would be confusing to do otherwise, given that Gell-Mann and Hartle's 
subsequent, and still tentative, modifications are motivated 
at least in part by the criticisms made here.
We will however comment briefly on the later version in footnotes.}

In part, though, we believe that Gell-Mann and Hartle's problems
arise from an attempt to maintain two irreconcilable lines of argument. 
As we have repeatedly stressed, a key question for an interpretation
of the consistent histories formalism is whether it does or does 
not {\it explain} our perception of a quasiclassical world.
It appears that  
in developing their program, Gell-Mann and Hartle had the problem of
the apparent persistence of quasiclassicality very much in mind.
They have introduced the notions of an information gathering and utilising
system, or IGUS,\refs{\gmhigus} and of a quasiclassical
domain,\refs{\gmhquasi} and they argue that certain
types of IGUS --- such as ourselves --- ``evolve to exploit'' the regularities
in some particular quasiclassical domain:\medskip\noindent
\medskip\noindent
``The one reason such systems as IGUSes exist, functioning in such a fashion,
is to be sought in their evolution in the universe.''
\refs{\gmhigus}\medskip\noindent
\medskip\indent
It is, they argue, as a consequence
of this that we perceive quasiclassicality as a persistent property of the
world around us, although, they claim,
nothing in the formalism ascribes a special
role to persistently quasiclassical consistent histories:\medskip\noindent
\medskip\noindent
``If there are many essentially inequivalent quasiclassical domains, then we
could adopt a subjective point of view ... and say the IGUS `chooses' its
coarse graining of histories and therefore `chooses' a particular
quasiclassical domain...  It would be better, however, to say that the IGUS
evolves to exploit a particular quasiclassical domain or set of domains.
Then IGUSes, including human beings, occupy no special place and play no
preferred role in the laws of physics.''\refs{\gmhigus}\medskip\noindent
\medskip\indent
This, Gell-Mann and Hartle maintain, is an argument which explains,
within the consistent histories formalism, the mystery
of why we continue to experience a 
quasiclassical world.  
It is an argument which, we suggest, to make sense, requires a 
particular, not at all straightforward, type of interpretation
of the formalism.  The same, we suggest, is true of Gell-Mann and
Hartle's discussions\refs{\gmhcomm, \gmhoverlap} of communication 
between IGUSes in incompatible quasiclassical domains 
or (as rephrased) of inference of each other's 
features.\refs{\gmhcommnew} 
We attempt to set out the sort of interpretation that is needed,
and the difficulties that it involves, in subsections 5.5.2 and 5.5.3. 
Gell-Mann and Hartle do not wish to advocate
such an interpretation.  While we sympathise with this view ---
since the interpretations described in 5.5.2 and 5.5.3 seem to 
us ad hoc, inelegant and convoluted --- we do not see how their
arguments can otherwise be supported. 

On the other hand, Gell-Mann and Hartle also maintain
that one can make {\it no} predictions within consistent 
histories that are not conditional on the set in which they are 
made.\refs{\hartlepriv, \gmhcommnew} 
We see this, in itself, as the core of a natural and internally 
consistent interpretation of the formalism --- which, for completeness,
we shall set out in Section 5.6.  
It leads, though, to the conclusion that the persistence of
quasiclassicality is an unexplainable mystery, and,
we suggest, contradicts the intent of
Gell-Mann and Hartle's extensive 
discussion of IGUSes and their evolution. 
It also entails a novel, and to us unacceptably ahistorical,
view of the past.
Hartle's statement that ``dinosaurs {\it did} roam the earth many millions
of years ago''\refs{\hartlepast} is {\it not} to be read as a 
statement about a
definite, observer-independent physical event; it is meant as a
shorthand for the claim that there is a particular consistent set
which includes the actual facts presented to us and in
which quasiclassical dinosaur density projection operators followed
largely
deterministic equations of roaming through the
Cretaceous era.
Other consistent fine grainings of the actual facts contain different
operators, incompatible with such a picture.
All these pictures are just theoretical constructs ---
a vast library of consistent historical fables, of which one is
free to choose any or none.\foot{``Fable'' is not, of course, 
Gell-Mann and Hartle's preferred term.} 
\foot{Whitaker,\refs{\whitaker} citing these comments of ours, puts it
well: in this view the formalism
might better be described as one of ``consistent stories'' rather
than ``consistent histories''.}  

We have found it impossible to reconcile these two positions, 
and feel that some clarification of Gell-Mann and Hartle's views 
would be useful.  There seem, however, to be divergent
views on what a clearly defined interpretation involves.
Hartle, for example, gives a revealing discussion
of what he calls the ``buzzwords'' of quantum mechanics in the appendix to his
Jerusalem lecture notes,\refs{\hartleone} and suggests:\medskip\noindent
\medskip\noindent
``Only a casual inspection of the literature reveals that many
interpreters of quantum mechanics who agree completely on the algorithms for
quantum mechanical prediction, disagree, often passionately, on the words with
which they describe those algorithms.  This is the `words problem' of quantum
mechanics.  The agreement on the algorithms for prediction suggests that
such disagreements may have as much to do with people as they do with physics.
This does not mean that such issues are unimportant because such
diverging attitudes may motivate different directions for further research.
However, it is important to distinguish such motivation from properties of
the theory as it now exists.''\medskip\noindent
\medskip\indent
Many physicists would certainly agree, and it is easy to understand
why --- interpretational arguments have often led nowhere, and real
progress has often been made by putting them aside and concentrating on
development of the formalism.
Still, the consistent histories formalism is a new description of quantum
theory, and its use in quantum cosmology does raise new questions,
in the light of which it seems sensible to reexamine the problem of
interpretation.

We persist in worrying about interpretations of consistent
histories, not only in the hope of motivating future developments,
but also because we do not
believe that there {\it is} full understanding of, or agreement on,
the algorithms for prediction.
For example, the interpretations we described in the preceding
subsections clearly do not predict, unconditionally, that we shall
observe a persistently quasiclassical domain.
Gell-Mann and Hartle argue that this can be explained, and we shall 
exhibit the implicit assumptions that we believe this argument
involves. 
Similarly, various interpretations differ on whether, and when,
communication between IGUSes is possible, and even on what is meant
by the word.
There seems to be no consensus on how to resolve all these
questions, and we do not believe this merely
reflects lack of thought or unfamiliarity
with the literature.  It seems to us, rather, that there are physical questions
which simply cannot unambiguously be answered by referring to the existing
literature, that the various possible answers rely on arguments which have
not always been set out in detail, and that it is necessary to try to
set out the possibilities.

The approach we have taken in this subsection 
is to look for a formal
interpretational scheme that can accomplish the aims set out in
Gell-Mann and Hartle's IGUS discussions, 
those of predicting our experience and explaining 
why it is of a quasiclassical world.
This divides into two parts.
We first examine the serious problems inherent in 
defining an IGUS.
Then, under the assumption that these problems can somehow be solved
or finessed, we set out some interpretations which are essentially
restrictions of the many-histories interpretation to some subclass of
the consistent sets.
We discuss Gell-Mann and Hartle's ideas within these interpretations,
and describe the new and rather awkward consequences which follow.
We stress, though, that these interpretations are entirely
unauthorised, and we expect that Gell-Mann and Hartle will reject
them.  
 
\vskip .2in
\leftline{\it 5.5.2 Problems of IGUS Definition}
\vskip .1in
We turn first to IGUSes.  The idea of an IGUS ---
a type of creature which is coupled by some form of sensory organs to its
environment, able to model the local environment by some form of
logical processing, and able to act on the results of its computations ---
is simple enough.
To characterise an IGUS within a particular consistent set it must, of
course, be identified with a coarse graining of that set,
just as any subsystem of the universe must.  The operators involved in the
coarse graining must possess a certain amount of temporal stability ---
they must be more or less the same operators at neighbouring times.
The functioning of the sensory organs of an IGUS, its logical processing,
and its resultant behaviour, can all then be described by correlations of the
coarse grained operators.

IGUSes then, are part of the formalism; had we wished,
we could have characterised supernovae in the same sort of way.
There is no significant notion of free will which can be attached to the
formalism. Just as in general relativity, IGUSes do what the equations say
they will do, though the predictions here are probabilistic.
Thus, if we study possible experiments which
might be performed by an IGUS, we must
bear in mind that it is the boundary conditions and probability rules
--- not any extraphysical free choice of an experimenter --- which
determines the experiment that actually takes place.
What any particular IGUS perceives can be bounded, though not completely
delineated, by the standard assumption of psychophysical
parallelism.\refs{\vonn}
That is,
the perceptions, sensations, thoughts --- in sum, the
consciousness --- of an IGUS must be paralleled in a natural way by
some description of the IGUS in the formalism.
Again, there is nothing unusual here; the same assumption is necessary to
make sense of any physical theory.

Now we come to an awkward question.  To set out any argument
about the experiences of quasiclassical IGUSes such as ourselves, and to
calculate any probabilities relating those experiences, we need to apply
psychophysical parallelism.  But to what?  Exactly what in the formalism is
supposed to correspond to a single IGUS?
One possible answer is suggested by the many-histories interpretation of the
previous subsection: each fundamental
consistent set contains a realised history, and
if an IGUS can be identified in the realised history, we can apply
psychophysical parallelism.
This is a perfectly clear, and quite natural,
prescription, but it is certainly not the prescription which Gell-Mann and
Hartle have in mind.  It tells us that if a consistent set contains a
quasiclassical description of an IGUS up to time $t$, and if --- once
again, following
Lemma 2 --- that set has a continuous family of extensions, then we must
apply psychophysical parallelism to each copy of the IGUS in each of the
extended sets.
We conclude that an IGUS just before time $t$, being unable
to tell which set to use, has no reason to expect quasiclassicality to
persist beyond time $t$, for there is clearly no reason
to suppose that psychophysical parallelism applied to a
non-quasiclassical description leads to quasiclassical
experiences.

In fact, Gell-Mann and Hartle are quite clear that they do not want
to speak of many copies of the same IGUS, associated with the various
extensions; in their interpretation of the formalism, there is only the one
IGUS.\refs{\hartlepriv}
They do not want IGUSes to correspond to a pair of objects:
a coarse grained history
and some particular consistent set in which it can be found.
Instead, they make the correspondence directly with the
coarse grained operators.\foot{More accurately, an IGUS should generally
be described by many coarse grained histories of non-zero
probability --- histories of zero probability may be ignored ---
within a consistent set,
since the course of its life need not generally be classically
determined.  We might, for example, arrange our travel plans to depend
on the results of quantum experiments.
A precise interpretation, capable
of supporting Gell-Mann and Hartle's arguments, would have to allow
for this. We shall ignore this complication, 
although it seems rather serious.}
However, once these 
coarse grained operators are identified, the remaining
details of the consistent set are not included in
the definition of the IGUS.  There are many consistent sets which
contain all of the coarse grained operators.  Any of these can be
used, but this is not supposed to imply a multiplicity of IGUSes 
corresponding to the sets.
It is further supposed that a quasiclassical IGUS will
continue to have the potential
for quasiclassical experiences so
long as there is some appropriate consistent set
which contains its quasiclassical coarse grained operators.

This, though, raises some new technical problems.
First, in defining an IGUS, one runs into the familiar problem
of where to make the cut between organism and environment.
Should humans be described only by the quasiclassical physics inside their
skins, or should a little of the environment be included so as to
allow a quasiclassical description of the information-gathering
process? Or should the relevant operators describe only the brain, or
some subset of brain activity?  No principle is known that would allow
us to identify any natural cut, and it is hard to see how one could
be found.  Of course, these problems would be relatively unimportant
if an IGUS were simply a rough characterisation of an
interesting class of objects within an already well-defined 
interpretation of the consistent histories 
formalism.\foot{This, in fact, is how Gell-Mann and Hartle's 
discussions of IGUSes are often understood.}  They are of crucial importance
here because, we claim, Gell-Mann and Hartle's arguments 
require an interpretation which uses the IGUS as a fundamental 
concept in its very definition. 
We shall give our attempts at such an interpretation in 
the next subsubsection.

A related point, to which we shall return, is that while it
is a plausible assumption that the experiences of an
IGUS are described in the same set as that of some coarse grained
operators which describe its information gathering and utilising
behaviour, it is still an assumption: in the end, it will need to be
justified by attaching a theory of experience to the formalism.

And then there is no reason to expect to find
a particular coarse graining which is uniquely
well-suited to describing the IGUS.  Which one should we use?  Why?
This raises the question of how we can tell
whether two IGUSes are the same. One
could imagine an IGUS such as the reader being described in terms of two
slightly different sets of operators, not coarse or fine grainings of each
other, not equivalent by our proposed equivalences. For example, one could
use the hydrodynamic variables defined by integrals of energy densities over
small spacetime regions marked out by some lattice which has the minimal
grid size to give decoherence, and also the same variables but defined with
the lattice translated by half a lattice spacing in some direction. We would
want to identify the two IGUSes in these two consistent sets and
give general conditions under which two descriptions are to be
identified as the same IGUS.

We can sharpen this last problem.
Since we want humans to be examples of IGUSes,
we must allow the operators to vary slightly over time in such
a way that after a long time they can be significantly altered.
This in itself poses no problem in describing human IGUSes, but it leaves
open the awkward hypothetical possibility that there exist two consistent
sets, one
containing an IGUS whose lifespan is described by some series of operators
and another, in which an IGUS is described identically to the first
for half its life and which then slowly diverges from the first, until the
operators which describe it are entirely different.
We have in mind here a picture in which
the operators from the two sets are inconsistent and yet
{\it both} sets of operators
describe the IGUS gathering and exploiting information from an
environment: nothing in the formalism or in the definition of an IGUS
implies that this is impossible.
If such an IGUS is calculating before the split, how is it
supposed to predict its future?
Which set of operators should it use?
Even a probabilistic answer to the last question would be satisfactory,
but the formalism gives none.
Still worse possibilities can be imagined, involving many branchings
and recombinings of the IGUS operators.

All of this suggests that, if
we are really to use the formalism to speak about
the experiences of general IGUSes then, even if we grant the
assumption that the experiences of the IGUS are closely linked to its
quasiclassical description, we need a much more precise notion of
how to define an IGUS than anything which has so far been suggested.
We need, in fact, a set of rules which tell us, a priori, which
sets of coarse grained operators describe IGUSes and which IGUSes are
distinct.  We can either, optimistically, hope that these rules can
somehow be formulated so as to
lead to no branching ambiguities of the type we have just mentioned,
or accept the ambiguities as possible features of the formalism; we
assume for the moment that the ambiguities can be removed.
There seems no prospect that any sensible, precise definition of an IGUS will
be found; still, let us assume there is one.
\vskip .2in
\leftline{\it 5.5.3 Interpretational Problems}
\vskip .1in
It is easy, given this assumption, to provide a 
formal, IGUS-centric interpretation.
Given a cosmological theory, we can identify all the consistent sets of 
coarse grained operators which correspond to (distinct) IGUSes in the
theory.  
Precisely one history is chosen from each of these sets, the choice being
governed by the standard probabilities, and the experiences
of the IGUS correspond to the history chosen from its set.
We are forced, in this interpretation,
to give up the hope of describing what happened in the very early universe,
or anything else which IGUSes do not experience.
In return we hope that we shall be able, for example,
to predict, {\it{unconditionally}}, that 
human IGUSes will continue to experience a quasiclassical world.
However, it will be apparent that this interpretation is solipsist.
For there is, in this 
interpretation, no correlation between the experiences of
these splendidly isolated IGUSes; each may well believe itself in
communication with others, but the others may be experiencing a quite
different history, or nothing at all.
Note also that by this device we have not
found a fundamental answer to the question of why
quasiclassicality seems to persist; we have simply shifted
the difficulty so that it becomes part of the IGUS definition problem.

Clearly we are not forced to use the definition of an IGUS in an
interpretation of this sort; one can try to eliminate the solipsism by
using something bigger.  The obvious choice is that of a quasiclassical
domain.\foot{Gell-Mann and Hartle's program aimed at
characterising the notion of a quasiclassical domain has sometimes,
understandably, been read\refs{\duerretal} as an attempt to define 
an interpretation of precisely this form.  This, though, is not
what Gell-Mann and Hartle intend: they do not, for example, regard
the question of whether a good characterisation exists as a test of
whether a sensible interpretation of the formalism can be defined.}
The same problems arise: quasiclassical domains might
branch; they are certainly imprecisely defined; we need a set of rules which
precisely characterises the coarse grained operators corresponding to a
particular domain.  Given such a set of rules, we can again follow the
same path: we use the probability weights to choose precisely one history
from each of the sets of coarse grained operators corresponding to
quasiclassical domains, and then suppose that each of these histories is
realised, and in particular
that each describes the experiences of all the IGUSes within the given domain.
This allows IGUSes in a quasiclassical domain to have genuine communication,
in which they can, for example, discuss features of the domain which form part
of their common experience.
Again, we have not explained our perception of persistent quasiclassicality
by this interpretation; we have simply assumed it.
Nonetheless, this seems more promising than the previous
interpretation: it avoids complete solipsism, and it is easier to believe that
quasiclassical domains can be characterised elegantly than 
that IGUSes can.

However, Gell-Mann and Hartle's recent discussion of
IGUS communication cannot be carried out in either of these
interpretations.  They envisage that:\refs{\gmhoverlap}\medskip\noindent
\medskip\noindent
``If two essentially distinct quasiclassical domains exist, they may overlap,
in the sense of having some features in common so that sets of histories
possess a common coarse graining.  It is then possible that IGUSes in
different but overlapping domains could make use of some of the common
features and thus communicate by observing and manipulating alternatives
of this common coarse graining.''\medskip\noindent
\medskip\indent
We have to be careful about the meaning of
communication in an interpretation of the consistent histories formalism.
It is important to distinguish between genuine
communication and the IGUSes' beliefs about the matter.
For instance, either of the interpretations
we have described allows a situation in which we can describe
two IGUSes --- perhaps both
working on the interpretation of the consistent histories formalism ---
each of which believes itself to be corresponding with the other,
but in which there is no agreement about the contents of the
correspondence.  No one, though, would describe this as communication.
We would like to
translate the statement that two IGUSes are communicating to mean that
the formalism allows us simultaneously to describe both their experiences,
including some correlated pieces of information, and gives a probability
distribution for the experiences which respects the correlations.
It seems to us that anything weaker can hardly be referred to as
communication.

Now, if the paragraph quoted is taken in anything
like the ordinary sense --- in which we live in and experience our
domain, and aliens live in and experience theirs; the formalism is
taken to mean that neither we nor they exist in multiple copies;
and we are understood to be transmitting and receiving
between the domains, and agreeing on the information
transferred --- then it is incorrect.\foot{
One can easily convince oneself of this by trying to describe the
communication process in a simple model.}
Probabilities of such a process may not be assigned in either of
the interpretations given above; neither, as we discuss below, can it
be done in any extended interpretation without some extra axiom.

Is there another interpretation in which we can understand the
paragraph?
We now investigate some possibilities.
Since the first interpretation above allows no communication
at all, and the second allows communication only within a quasiclassical
domain, they cannot support Gell-Mann and Hartle's
conclusion.
So, why not adopt the obvious solution, and make a similar interpretation
using something other than quasiclassical domains?
The problem is --- what?  If a pair of IGUSes $I_1$ and $I_2$
live in distinct quasiclassical
domains $D_1$ and $D_2$, and if there is some consistent
set $\S_1$ containing a description of both IGUSes (at least for some
period of time), one
can always declare by fiat that one history from $\S_1$ is realised, in which
case communication between $I_1$ and $I_2$ is restored.
But how, abstractly, is $\S_1$ to be characterised?   And what are we to do if
there is another IGUS $I_3$ in a third domain $D_3$, and
sets $\S_2$ (containing a description of the
IGUSes $I_1$ and $I_3$) and $\S_3$ (containing
$I_2$ and $I_3$), with the property that $\S_1$, $\S_2$ and $\S_3$ 
are pairwise
inconsistent?  If $I_1$ and $I_2$ are in communication,
then so must each pair of IGUSes be.
Yet the formalism gives no
joint probability distribution for the experiences of the three and,
even if we wished to go beyond the formalism, Lemma 8 tells us that
we cannot generally find {\it any} joint probability distribution consistent
with those describing the pairwise communications: in other words,
the probabilities specified by the decoherence functional distribution
on the sets $S_i$ are incompatible.

The question is clear: either $I_1$ and $I_2$ are in communication, or they
are not.  In particular, we ought to
be able to derive a probability for $I_1$---$I_2$ communication, and
similarly for the other pairs.
We cannot get around this by
suggesting that $I_1$ has a choice of communication with $I_2$ or $I_3$, but
not both, since the formalism allows no notion of free choice: however things
seem to $I_1$, it must be possible to calculate the probabilities of the
decisions $I_1$ appears to make.
Nor can we exclude the possibility of communication between any of the pairs,
without either excluding the possibility of communication between
all distinct quasiclassical domains, or else finding some rule which explains
under which circumstances IGUSes in distinct domains can communicate and
which gives a joint probability distribution for the experiences of all
IGUSes in communication.
The first of these would contradict Gell-Mann and Hartle's discussion;
the second would go beyond the formalism.

Here we have used only the assumption that, if a pair of IGUSes are in
communication, there must be a joint probability distribution for their
experiences,
which we believe must be true under any standard definition
of communication.
We have {\it not} explicitly assumed that pairwise communication implies that
information can freely be shared among the three IGUSes; we note, though,
that the literature again gives no clear and generally agreed rule
which either forbids this or imposes any other well-defined constraint on
the information which an IGUS can pass on.

We must conclude, then, that Gell-Mann and Hartle's recent discussion 
of interdomain communication is wrong.\foot{
We should 
mention here that the current draft of the paper \refs{\gmhcommnew}
differs from the preprinted version quoted from in the text in that 
it contains no explicit reference to ``communication'' between 
IGUSes. We believe however that our criticisms are just as 
relevant to the current draft.  The 
claims remain virtually unaltered:\refs{\gmhcommnew}\smallskip\noindent
`` It is...possible that IGUSes evolving in different but overlapping
realms could make use of some of the common features ... to infer
features of each other. The problem of inferring the existence
of other IGUSes using distinct but overlapping realms is not so very
different from that involved in ordinary searches for extraterrestrial
life''\smallskip\noindent
That it {\it is} rather different, we believe we have demonstrated in the 
text. The IGUSes are now inferring the existence
of each other rather that communicating, but our arguments 
depended only on the translation 
that there exist a joint probability
distribution  for the IGUSes and so apply equally well.}

Since this is a strong claim, let us repeat the main argument. 
We proposed two interpretations in which the fundamental notions
are, respectively, IGUSes and quasiclassical domains. In the first,
there is no communication at all; in the second there is communication
only within quasiclassical domains.  Consider then the 
naturally extended 
interpretation in which two IGUSes are in communication 
when there is a consistent set that includes them both.
This interpretation
leads to a contradiction when we have three IGUSes, pairwise 
in communication, 
for which there exists no joint probability distribution
on all three 
which respects the joint distributions for the pairs.
The interpretation demands that a given IGUS concludes that it is 
communicating with the other two, but this contradicts the lack
of an overall joint probability distribution.

What of Gell-Mann and Hartle's discussions in which IGUSes ``evolve
to exploit'' certain variables?  
This is the argument IGUS-centric interpretations are designed to
support.  Given a clear definition of IGUSes, one could 
indeed hope to study their general properties, observe that
IGUSes do tend, as time progresses, to become described by
operators which correlate better and better with more and
more quasiclassical variables and which perform more and
more sophisticated calculations simulating
quasiclassical equations of motion.  
And, given the assumption that IGUSes are {\it the} fundamental
objects in the theory, one could then use these results 
(together, as always, with the assumption of psychophysical
parallelism) as an {\it explanation} of their (and in particular
our) perception of a persistently quasiclassical world. 

In an 
interpretation in which quasiclassical domains, rather than
IGUSes, are fundamental, IGUS evolution can be described 
and investigated. 
The only variables that IGUSes can 
{\it possibly} be described as exploiting, however, 
are quasiclassical, so that one can hardly speak of 
IGUSes as {\it evolving} so as to exploit these variables rather than
others.  
Since it is in this sense that 
Gell-Mann and Hartle's statements about evolution are to be
understood, this interpretation is inadequate to accommodate them.
And, of course, in this 
interpretation, the persistence of quasiclassicality
is imposed by fiat rather than explained. 

In the remainder of the subsection we shall reiterate the 
properties of the IGUS-centric
interpretation which is the closest we 
have been able to come to the spirit of Gell-Mann and Hartle's
ideas of explaining our experience (though as we have seen 
it fails to allow some of their more detailed assertions).  

We can --- it is hoped --- predict what
we shall experience by identifying the relevant set of future
projection operators, and then test these predictions.
We have mentioned the problem of branching ambiguities.
Perhaps there is no general way of finding these experiential projection
operators for arbitrary IGUSes, for human IGUSes in arbitrary
cosmological models, or even for some future humans in our own
cosmology.
But in defence it might be argued that none is
needed: any IGUSes afflicted by branching ambiguities
will (or may) experience something after the branching;
even if they have perfect knowledge of the boundary conditions they cannot
make any probabilistic statement about what (or whether) they
will experience.  This would be scientifically unfortunate for 
them, but need it concern us?  
Should the hypothetical problems of
other IGUSes, or even the real problems of humans in some future
circumstances, be a deadly objection to a scientific theory?
If, at present, we can predict our own future experiences, should
we not be content?
 
Possibly, but this still
begs the question: {\it can} we identify the operators that allow
us to predict our own future experiences?  Nobody seriously
claims that the operators can be precisely identified at the moment.
But many claim that we can say roughly what form the operators must
take: that they are almost surely quasiclassical descriptions of the
brain in operation.  It is important here to make the distinction
between the practical question (what is the state of things?) and the
theoretical question (do we understand why?).  On the practical
question, we agree: we have no reason to suppose that anything
other than a quasiclassical description of the brain is needed.
But we see the theoretical question as
crucial.  At a fundamental level, we do not
understand why
the operators relevant for describing our
experience  must be a subset of
the quasiclassical projection
operators which are most convenient for describing the brain's logical
functioning.
It is an assumption --- a necessary assumption to make
the formalism describe our experience; a plausible assumption as part
of a general theory of experience; but still, no more than an
assumption.
This forces the conclusion that we cannot answer the question as to why
we perceive a persistently quasiclassical world within the formalism.
Instead, this question becomes part of the general problem of finding a
theory of experience.  And to do the job we require --- to explain,
finally, why we perceive a quasiclassical world --- the sought-for theory of
experience will have to be framed not in quasiclassical terms (which
would merely assume the answer) but in the language of the consistent
histories formalism.

At the risk of repetition, let us stress this crucial point.  To do
the job, a theory of experience would have to take as input some
characterisation of an IGUS and return as output a list of the projection
operators describing its experiences.  There is no {\it a priori} link
between the behaviour of the IGUS, as described in some history of
some consistent set, and its experience.  Indeed, it is crucial that
the link is generally absent, since we can be described (albeit
perhaps inconveniently) in many consistent sets which clearly are not
suitable for describing our experiences.  Hence the theory would have,
for each IGUS, to construct such a link with a particular set and explain,
from its axioms, why this set is the correct one.

Nothing in the consistent histories account of our evolution implies
the form of such a theory, and no coherent theory of experience, either
quasiclassical or quantum theoretic, has yet been framed.
There is no consensus on the form such a theory might take and, indeed, some
doubt that a scientific theory of experience can ever be found.
Nor is there any evidence that any such theory should
be naturally framed in the language of the consistent histories
formalism.\foot{Moreover, if a theory really can be found to identify the
scientifically relevant set or sets
of histories, perhaps the consistency criteria
need not separately be imposed.  For one can certainly imagine the
consistency of the scientifically relevant sets following directly from the
hypothetical theory.  In this case there seems no need to discuss
general consistent sets at all.}
For consistent historians to appeal to an unknown theory of experience
in order to make unconditional predictions is then, at best, to
postpone the question of whether such predictions can in fact be
made.

Another concern is the solipsism which
the IGUS-centric interpretation implies.  We can describe our
own experiences within the formalism.
We can also describe other IGUSes', but not
simultaneously.

 We cannot, moreover, see how to
avoid solipsism in extended interpretations of this
type: to avoid it one needs to ascribe reality to at least one 
set that extends our experience. If one ascribes reality to all 
such sets one then runs into the difficulties encountered in our
discussion of the many-histories interpretation: when these histories
contain quasiclassical descriptions of IGUSes, these descriptions
generically cease to be quasiclassical very abruptly. So one requires
a selection principle as to which sets to ascribe reality to --- 
quasiclassical domains are a possibility we have mentioned 
but this,
again, has its own problems.

Neither of the features to which we have drawn attention
is unprecedented.
It is disappointing that the problem of the
apparent persistence of quasiclassicality is left as a mystery to be
solved by some future theory of experience, but other approaches
to quantum mechanics use the same strategy, which goes back at least
to Wigner.\refs{\wigner}  It is awkward that experience thereby becomes
entangled with the quantum formalism at a fundamental level, but of
course this could conceivably turn out to be unavoidable 
--- who can tell for sure?
Likewise, it is not unknown for interpreters of
quantum mechanics to find themselves driven to solipsism,
and some physicists find this a respectable scientific
position.\foot{We do not, but this debate goes beyond our scope.}

Gell-Mann and Hartle characterise their program as essentially
an elucidation of quantum theory. Indeed they often 
refer to the consistent histories approach as ``quantum mechanics''
or ``quantum theory.'' We would quarrel with this nomenclature
since the consistent histories is only one among many, 
currently existing and potential, descendants of
the quantum mechanics of Bohr and Dirac. 
We do, however,  agree with them that  
it is a particularly important development.
One can trace a line of thinking in which, gradually,
through the work of many people, the provisional or irrelevant
notions cluttering certain earlier interpretations have
been stripped away.
No notion of measurement by classical apparatus is necessary; no
collapse of the wave function need be considered --- this much was
suggested by Everett.  Everett's attempt at a many-worlds
interpretation, though, was left incoherent by the lack of
any criterion by which physical histories could be extracted from
the formalism.
Griffiths, Omn\`es, Gell-Mann and Hartle have supplied such criteria.
We are left --- and we think this is most clearly seen in the quantum
cosmological picture of Gell-Mann and Hartle --- with a clear
characterisation of the fundamental points on which there
is disagreement about the interpretation of standard quantum
theory.
As Penrose has stressed in other contexts,\refs{\penrose} there is
a fundamental divergence between those who seek to add a theory
of experience to quantum theory, and those who would prefer a
theory of reality.
Perhaps the debate really has finally reached impasse, and can only be
resolved by new science on one side or the other.
Certainly, each is left with a hard task: one side needs to
construct the theory of experience which will shore
up their position; the other to find a realistic theory which replaces or
extends quantum theory.

One can imagine a variety of radically different outcomes to these programs.
But, if the consistent histories formalism is fundamentally the right
setting, then each program could be completed by a selection 
criterion which
picks out particular consistent sets:\foot{Such a criterion need not
be deterministic: in principle, a probabilistic criterion defined 
by a measure on the space of consistent sets could do the job ---
if a sensible measure could be found.} on the one side, a class of sets
would be selected to describe the experiences of IGUSes, or at least
humans; on the other, a single set would be used to describe reality.
It is common for those who appeal to a theory
of experience to claim that, while alternatives are not 
excluded, theirs is the natural null hypothesis.
In the light of the consistent histories formalism, and in particular
of these last observations, we suggest that this position is very clearly
untenable.

\subsec{The Unknown Set Interpretation} 

Most of the interpretational ideas in the 
literature add complicated new ideas (``truth'', Logic, and the
IGUS, to name but three)  without, it seems to us, resolving any of 
the problems in making sense of the consistent histories formalism.  
In this subsection we suggest a simpler interpretation, 
which we think achieves all that any other interpretation has
achieved without adding conceptual frills or suggesting a 
resolution of unresolved problems.  
It is simply this: the world is described by precisely one history
from one consistent set.  Given the set, the history is chosen 
randomly according to the decoherence functional probabilities.
We do not know which is the correct set, or how it should be 
characterised, or why it has the properties that it appears to have.
If we are willing to assume an agreed list of 
historical events (in the 
terminology we have adopted from Omn\`es, actual facts), we can 
pin down some of the past sets of projections in the realised history.
These, however, as we have seen, will not generally allow us to make
useful unconditional predictions.  

What can we make of the apparent persistence of quasiclassicality
in this interpretation?  It is either an illusion or a mystery --- 
we have the impression that the realised history has been 
quasiclassical so far, 
and that is all there is to be said.  
As a practical matter, we should no doubt assume that the realised
history will be quasiclassical in the future.  
Conditioned on this assumption,
we can make probabilistic predictions with the same scientific 
content as those of the Copenhagen interpretation. 
Then, if we are to assume persistent future quasiclassicality, we may as
well (since it hardly deepens the mystery further)
assume past quasiclassicality too.  
However, this interpretation does not pretend any explanation of these
facts.  

This interpretation, it seems to us, is in practical terms
equivalent to Griffiths' Logical interpretation, to the 
reasoning suggested by one of Gell-Mann and Hartle's lines of
argument (though not their discussions of IGUSes), and to Omn\`es'
approach (unless and until any interesting new definition of
future true facts can be found).  
It has the advantage of clarity: indeed, it seems to us the 
clearest formulation of quantum mechanics, and the best understanding
of that theory currently attainable, that involves neither
a set selection hypothesis nor auxiliary variables.  Either of
these could in principle solve the problem it leaves open --- the
mystery of persistent quasiclassicality.  

We have come full circle.  
As we have said already in our discussion of Omn\`es interpretation,
our own view is that persistent quasiclassicality is such a basic 
feature of our
experience, and failure to predict it leaves such a significant gap in
our understanding of the world, that we feel that some argument, or
some new interpretation, or some new formalism, that would amend the situation
should be sought.  
Others may find the present interpretation perfectly acceptable as
a fundamental theory.  We hope, at least, that it will be recognised
that it {\it does} leave a mystery which {\it could} possibly have 
an explanation. 

\newsec{Persistence and Prediction}

Much of our discussion on formal interpretations has involved the question
of whether our persisting experience of a quasiclassical world can be
predicted within a given interpretation.  It seems to us {\it the} crucial
question.  If it were not for their failure on this score, both
the Omn\`es and the many-histories interpretation would be quite acceptable;
if Gell-Mann and Hartle's arguments, which lead to the conclusion that
our experience {\it can} be predicted, did not rely crucially on premises
about human experience, they would be complete.
It also seems to us a perfectly legitimate scientific question, and we would
no more happily accept a fundamental theory which cannot supply an answer
than we would accept one which cannot explain celestial mechanics.
Still, it is not a traditional question, and this may lead some to
conclude that it is somehow beyond the scope of science, and not something
about which down-to-earth equation-solvers need be concerned.

However, a little reflection shows that this is false.
There are obvious quantum
cosmological effects --- the tunnelling through to a region of true vacuum,
propagating at the speed of light, for example --- which would destroy
(or at least discontinuously disrupt) persistent quasiclassicality.
If a theory predicts that a vacuum bubble should engulf the solar system
at some time after noon, Poisson-distributed with mean 1 hour, then,
should we survive till midnight, we will certainly reject the theory.

Moreover, once non-trivial final
density matrices are allowed, we need not rely on manufactured cosmological
catastrophes to cause the breakdown of quasiclassicality.
Quite the converse: the inevitable, consistent-set-independent breakdown
of quasiclassicality becomes a generic feature.
Suppose that we are given any model in which some consistent
set $\S = ( \rho_i , \{ \sigma_1 , \sigma_2 , \ldots \},
\{ t_1 , t_2 , \ldots \} )$ describes a persisting quasiclassical domain.
(We have included the projection times, since they are necessary to
establish quasiclassicality.)
Choose any time $t$ such that there is some $r$ with
$t_r < t < t_{r+1}$ and such that the truncated set
$\S_t = ( \rho_i , \{ \sigma_1 , \sigma_2 , \ldots , \sigma_r \},
\{ t_1 , t_2 , \ldots , t_r\} ) $ is non-trivially extended by
$ \sigma_{r+1}$.
Then Lemma 4 implies that we can
choose a final density matrix $\rho_f$ such that
$\S_t^f = ( \rho_i , \rho_f , \{ \sigma_1 , \sigma_2 , \ldots , \sigma_r \},
\{ t_1 , t_2 , \ldots , t_r\} )$
is consistent and that the corresponding histories in $\S_t^f$ and $\S_t $
have the same probabilities, but such
that $\S_f = ( \rho_i , \rho_f , \{ \sigma_1 , \sigma_2 , \ldots ,
\sigma_{r+1} \} , \{ t_1 , t_2 , \ldots , t_r , t_{r+1} \} )$ is
inconsistent.

In other words, if we have a theory describing our own quasiclassical domain
up to the present time, and if the theory tells us that the results of some
quantum experiment we are about to perform are genuinely unpredictable,
then we can find another theory which reproduces the description up to the
present time, but in which the standard quasiclassical description of the
experimental results cannot be made.
This, at least, is true if the original theory had pure initial density matrix
and trivial final density matrix, since under these conditions Lemma 4 holds
in both finite-dimensional and infinite-dimensional Hilbert
spaces.\foot{No doubt similar results hold far more generally.}

Clearly, then, we cannot avoid discussions about the
breakdown of quasiclassicality, and
its implications for the experience of IGUSes such as
ourselves: if we want to understand general quantum cosmologies, then
the formalism forces these questions on us.  It seems, in fact, that among
theories accurately describing the present quasiclassical world, nearly
all will predict the eventual breakdown of quasiclassicality.
In particular, any interpretation that discusses the experiences of IGUSes
has to assume something about the experience of an
IGUS whose lifespan is not describable quasiclassically beyond a certain
point in time.  This is quite distinct from any
conventional picture involving the quasiclassical destruction of the
IGUS.  No doubt the simplest assumption is that any experience requires a
persistent quasiclassical description.  But again, while this is a
useful assumption if one wants to avoid awkward questions about the
predictability of future experience, it is still an assumption.
If one makes it, one must suppose that it will somehow follow from a
general theory of experience, and this is pure guesswork at the
moment.  Even if we assume that when there are
relevant quasiclassical variables they invariably correspond to
quasiclassical experience, we can deduce nothing about the situation
when there are no relevant quasiclassical variables.

It is also interesting to note that Lemma 4
implies that we cannot make any useful predictions, either unconditional or
conditional, without a theory of the final density matrix.
Neither consistency with the actual data, nor the assignment of a
particular probability to those data, can be used as selection
criteria for $\rho_f$: at any given point in time, a large
family of final density matrices will do equally well.
This does not appear to be a grave problem:
there are apparently sensible
choices --- such as taking $\rho_f$ to be $I$ ---
which can be made on grounds of simplicity.

\newsec{Why be consistent?}

Since we consider in this paper several different
versions and interpretations of the consistent
histories formalism, we here discuss why any consistency criterion
should be retained.
We first mention recent interesting work by Goldstein and
Page,\refs{\goldsteinpage} who use an earlier result of
Page \refs{\page}
to show that the
Gell-Mann and Hartle consistency conditions can be considerably weakened while
preserving the probability sum rules.
Goldstein and Page point out that if
${\cal {S}} = (\rho_i , \rho_f , \{\sigma_j\}, )$ is a consistent set
according to \decohgmh, then
\eqn\pg{\eqalign{
\Tr ( \rho_f P^{(i_n )}_n  \ldots P^{(i_1 )}_1
   \rho_i P^{(i_1 )}_1 \ldots P^{(i_n )}_n ) & =
\sum_{i'_1 \ldots i'_n}
\Tr ( \rho_f P^{(i'_n )}_n  \ldots P^{(i'_1 )}_1
   \rho_i P^{(i_1 )}_1 \ldots P^{(i_n )}_n ) \cr
& = \Tr ( \rho_f \rho_i P^{(i_1 )}_1 \ldots P^{(i_n )}_n ) \, ,}}
giving a simpler expression than \probtwo\ for probabilities:
\eqn\probthree{
p( i_1 \ldots i_n ) = \Tr ( \rho_f \rho_i P^{(i_1 )}_1
\ldots P^{(i_n )}_n ) \, .}
They go on to observe that, since \probthree\ is linear, it
automatically satisfies the sum rules \sumrules\ so long as the
projections $P^{(i_r )}_r$ belong to projective decompositions
$\sigma_r$, whether or not they form a consistent set.
For general sets of $\sigma_r$, the
expressions for $p( i_1 \ldots i_n )$ include negative
quantities, and so cannot be given a probability interpretation.
However, one can simply restrict to the sets for which all
$p( i_1 \ldots i_n )$ are positive.
Goldstein and Page refer to these as sets of linearly positive
histories.  As they point out, these sets have all the relevant
mathematical properties of consistent sets, and give a
broad extension of the consistent histories formalism.

Goldstein and Page further point out that if one uses their conditions
there is no obvious mathematical reason to require $\rho_i$
(or indeed $\rho_f$) to be
positive semidefinite: whether they are or not, one can
simply restrict attention to the sets whose histories have positive
probabilities.\foot{This loss of structure seems a little worrying, and
suggests that the Goldstein-Page conditions, taken alone, are rather
too loose: as Goldstein and Page themselves stress, further
selection principles are needed.
We prefer to consider the Gell-Mann--Hartle conditions for the
pragmatic reason that they produce fewer consistent sets.
However, as Goldstein\refs{\goldsteinpriv} points out, even here
things are complicated: it is not so clear that the Gell-Mann--Hartle
conditions necessarily
produce fewer fully fine grained or maximally refined sets.}
We include linear positivity amongst the class of possible
consistency conditions when we refer below to consistent sets.

Now let us consider a more general possibility.
We have seen that if a set $\S$ satisfies any of the various
consistency conditions
then its histories obey the standard probability sum rules.  This is
certainly a very pleasing property, and gives an interesting selection
criterion for sets of projective decompositions.  But are there compelling
reasons for restricting attention to such sets?  After all, even for
a general set $\S$ for which either the initial or final
density matrix is unity, the weights \probtwo\ still have the fundamental
properties of a probability distribution:
\eqn\probfour{
p(i_1 , \ldots , i_n ) \geq 0 {\rm~and~}
\sum_{i_1 , \ldots, i_n} p(i_1 , \ldots , i_n ) = 1 \, . }
Even if both $\rho_i$ and $\rho_f$ are non-trivial, one can in general
simply renormalize the individual weights, which are all non-negative, so
that their sum is one: this fails only if all the weights are zero.
Why not simply accept these
weights as probabilities for the
histories, and calculate those for coarser-grained histories
simply by summing: why not, in other words, accept the right hand side
of \sumrules\ as the correct coarse grained probability calculation, and
ignore the left hand side entirely?  Is it possible to make physical sense of
such probabilities?

This last question was raised in Griffiths' original
article:\refs{\griffincon}\medskip\noindent
\medskip\noindent
 ``Another direction in which one might hope to extend the consistent
histories approach is to find some physical interpretation of the
weights $\ldots$ when the events in question do {\it not} form a
consistent history. One such interpretation is already implicit in
Theorem 7 of section 5: aside from normalisation, the weights for
histories belonging to a particular
(inconsistent) family are the probabilities that the corresponding
{\it consistent} histories {\it would}
occur in a combined system which includes idealized measuring instruments
which detect the different events in the original system at the appropriate
times.  However, this interpretation is neither simple nor a source
of much intuition, given all the peculiarities associated with
quantum measurements.  Can one do better?''\medskip\noindent
\medskip\indent
One can: it is perfectly possible to modify each of the
interpretations outlined in section 6 to allow for inconsistent sets.
This causes no logical
contradiction: it is trivial to extend the many-histories
interpretation to inconsistent sets, and with a little more care the
other interpretations also extend.

An immediate objection is that this weakens the formalism
enormously, since now even the actual facts barely test a cosmological
theory: any collection of
actual facts can be incorporated in many different inconsistent sets, and we
cannot even make use of probabilistic tests, since the
probability assigned to the actual facts depends on the inconsistent
set in which they are embedded.
This would be the end of the story, were it not
for the point which we have stressed --- 
any satisfactory interpretation of the consistent
histories formalism as a fundamental physical theory 
must eventually identify one particular consistent set in 
which to calculate.\foot{
As we have seen, there are several different lines of argument which
could be used to justify different possible choices.} 
This raises the question as to why we should insist that the choice
be restricted to consistent sets.  Could a theory of experience
perhaps rely on a correspondence of mental states with
operators in an inconsistent set?  Or could those who would plump 
for a set selection criterion instead  
postulate that reality is described by some
particular inconsistent set?

It has to be admitted that these are possibilities.  There is no
logical contradiction involved in using an inconsistent set in this
way.  The reason why we tend to reject these proposals is that
inconsistent histories disagree with experimental data, unless they
are fine tuned to be undetectably inconsistent.
Thus the success of conventional quantum mechanics, which tells us
that experimentally observed conditional
probabilities relating various events can
be calculated using only an initial density matrix, the hamiltonian, and
a description of those events in the language of projection operators,
is strong evidence against generic inconsistent histories.
We know, in practice, that we are able to ignore the parts
of the universe outside the spacetime region of our experiments and
to coarse grain the description of the experiment itself, and
the consistent histories formalism guarantees that this
coarse graining works.
If nature were really described by
a history from a generic inconsistent set, with the probability interpretation
above, then the successes of
reductionist science and the reproducibility of simple experiments
would be inexplicably lucky accidents.
For example, to calculate the rate of a chemical reaction in a generic
inconsistent set, one would have to calculate the probability of every
history in that set.  Depending on the projections in the set, this could
involve complicated calculations of the nuclear spin trajectories
of the atoms involved, or of the local matter densities in distant nebulae,
or of many grotesquely uninterpretable variables.  There would be no
reason to expect that the calculation for any pair of atoms could be
performed according to Copenhagen (or any other simple) rules of
thumb and, even if such a calculation could be performed,
there would be no useful connection between the results for different
pairs.
In contrast, if the
set were consistent, one could calculate the rate by considering only
the coarse grained operators corresponding to various chemical densities
inside the test tube and then applying the usual Copenhagen rules.

An inconsistent histories formalism is not absolutely excluded.
It is still possible that the universe is described by an inconsistent set,
the probabilities of whose histories, when summed to give the probabilities
of events such as the measurement of the position of an electron on
a screen in a double slit experiment, give the observed answers.
If a theorist produces an otherwise elegant
and successful quantum cosmological
theory which implies that nature is described by a particular inconsistent
set, we should want a very good explanation for the apparent consistency
(in both senses) of our experimental data, but we should not reject the
theory out of hand.
However, assuming that one uses the decoherence functional to define
probabilities for coarse grained histories, it seems to us
that presently the case for some
consistency criterion in quantum mechanics is compelling.

This, it should be stressed, is 
not a criticism of other history-like formulations of
quantum mechanics,\refs{\bohm,\ \samols,\ \sorkin} or of related
theories.\refs{\grw}
There are sensible probability measures apart from the one
defined by the decoherence functional
and sensible sample
spaces apart from spaces of coarse grained histories,
and once one abandons one or both 
the comments above no longer apply.

\newsec{Conclusions}

The virtues of the consistent histories approach are worth
reasserting.
We have a natural mathematical criterion which is empirically
supported and which identifies the physical content of quantum
theory to be propositions about particular collections of events
encoded in the consistent sets.  This gives a framework in which
attempts to set out a quantum theory of the universe, and the problems
inherent in this idea, can be sensibly discussed,
and in which intrinsically quantum cosmological questions (such
as the role of early anisotropies in seeding galaxy formation) can,
at least in principle, be posed.
There is a natural and attractive path integral version of the
formalism, which is manifestly Lorentz invariant.  The formalism
gives a new way of thinking about the problems of quantum gravity, and
of studying measures of information in quantum theory.
And, as Gell-Mann and Hartle have stressed,
the abstract characterisation of a quasiclassical domain within the formalism
is a new and interesting research program.
Many attempts have been made to find natural mathematical structures
and interpretational postulates that allow one to use quantum theory 
to make statements about 
physics that go beyond the Copenhagen bounds. 
In our view, the consistent histories formalism is one of the most
significant developments.  Whether it will eventually form part
of a theory with greater explanatory power than the Copenhagen
interpretation remains in doubt --- the possibility cannot be excluded,
but there is, as yet, no positive
evidence. 
Meanwhile, the consistent histories approach has the undoubted
virtue of illustrating, more sharply than its predecessors, the
problems inherent in quantum theory.

Advocates of the formalism thus have a good case.  Perhaps we have
even made one or two marginal additions.  The suggestion that
approximately consistent histories might play a key role
has worried many who, like us, prefer the equations of their
fundamental theories to hold exactly: it is reassuring to note that
any interesting physical process can almost certainly be characterised
by exactly consistent sets.  The derivation of consistency criteria
from the probability sum rules is an absolutely key insight, and is
obviously mathematically very natural, but leaves some wondering whether the
criteria are justified by any scientific principle.  The answer is
clear, and has been repeatedly pointed out by the authors of the
formalism by illustrating the failure of inconsistent histories to
reproduce the observed data in two-slit and other quantum experiments.
But since obstinate critics can always say that such illustrations only exclude
very particular inconsistent histories, a general discussion of
inconsistency and reductionism may be helpful.

Nonetheless, we feel that the worrying features of the consistent
histories approach deserve more attention.
Firstly, there are important points of principle to be resolved:
at present, there is no clear agreement on how the actual facts should be
selected, nor on which criterion (if any) should be used to deduce
properties beyond the actual facts.
Secondly, as we have seen, there are several
possible attitudes which consistent historians can adopt to the past,
and it would be good to see this question addressed more clearly.
If one accepts the formalism as it stands, and introduces no
historical actual facts, one seems to be led towards
solipsism of the present, since it is unlikely that any past event can
be described in every consistent set, and each consistent set gives an
equally valid description of the past.
Similarly, if Omn\`es' criterion for true propositions is adopted ---
and no other inferential rule has been suggested in the
literature to date --- the formalism almost certainly allows no ordinary
quasiclassical deductions to be made unambiguously.
It seems clear, practically speaking, that we cannot deduce, from the
tides, from our perception of moonlight, or from any quasiclassical event
on earth, that the moon is in a quasiclassical orbit.
And, as we have argued, although it is hard to find a rigorous
demonstration, this does also seem likely to be a problem in principle:
a quasiclassical description of events on earth should be consistent with
an infinite number of pictures, in nearly all of which the moon does
not behave quasiclassically.
Since the most natural collection of actual facts appears to be our
own experiences, and since the same inferential problem then arises
for statements about our fellow creatures, the formalism also
appears to lead to personal solipsism.

The prediction of the future is an even more pressing difficulty.
Why do we continue to experience a quasiclassical world?
The only answer we have found in the literature
is that supplied by Gell-Mann and
Hartle.  We predict that we will experience a quasiclassical
world because our experience will be described by certain decohering variables,
which, for now, we simply assume to be quasiclassical ones coupled to
a quasiclassical environment, and
which a to-be-found theory of experience will identify as the
fundamentally correct ones amongst all the possibilities offered by
the formalism.
We must assume that such a theory of experience can be found, since we
{\it do} experience a quasiclassical world.
We put the argument starkly, but we believe the translation is
accurate.  It is a coherent position, and anyone who holds it
will feel happy developing the formalism along the current lines.
No one, though, will mistake it for an ultimately satisfactory answer
to the question.  If it is correct, there seems to be no hope of our
understanding the answer in the foreseeable future.  We are clearly
a very long way from a theory of experience, and there is no guarantee
that such a theory, of the type needed, will or can ever be found.

There is an alternative which cuts through all these problems.  It is
to accept, once and for all, that quantum theory is not sufficient
to describe the world, and that it should be augmented by a further
axiom which takes the form of a selection principle.
The consistent histories formalism has taught us that there are
infinitely many incompatible descriptions of the world within quantum
mechanics.  Perhaps some simple criterion can be found to pick out one
of these descriptions, by selecting one particular consistent set.
Such a criterion should explain persistent quasiclassicality, not as a
consequence of our own biased perceptions, but as a deducible fact;
it should remove all solipsist tendencies from the theory;
it would restore definiteness to statements about the past.
To be remotely persuasive, of course, any such criterion would have to
be simple and elegant: there is no point in merely setting out a
characterisation of the world as we see it.

Consistent historians will certainly agree that this is an interesting
program.  Omn\`es has recently set out a speculative proposal
along precisely these lines.\refs{\omnesnew}
As others have noted,\refs{\duerretal , \hartlepriv}
Gell-Mann and Hartle's program aimed at characterising the
notion of a quasiclassical domain could equally be viewed
as a search for a set selection criterion, though this is
not their motivation.
Moreover, Gell-Mann and Hartle's explorations of stronger decoherence
criteria (such as $M$-decoherence) and of alternative set
selection rules (such as replacing projections by sums over
fine grained position space histories) point out ways in which
the set selection problem can, at least, be diminished.

Nonetheless, the search for a selection principle which will pick out
a unique consistent set is a line of development that 
tends to be placed firmly in the category of unorthodox
proposals occupied by alternative dynamical theories such as
those of Ghirardi-Rimini-Weber\refs{\grw} or Gisin\refs{\gisin} and
Percival\refs{\percival} and by auxiliary  
variables theories.\refs{\bohm,\ \samols}
Interpretations of the consistent histories formalism which
do not explicitly rely on set selection,
on the other hand, have been regarded as setting out
{\it the} correct development of
quantum theory and defining the natural null hypothesis given the
present experimental data.
If there is one single point which we would wish to emerge from
this paper, it is that this view is indefensible.
The apparent persistence of quasiclassicality is a
central problem for the consistent historians.  Either it is
fundamentally bound up with the problem of a theory of experience, or
the two problems are separate.  In the former case, there is presently
no hope of a solution.  In the latter case, there are many interesting ideas
which can be explored.  In the former case, we are led
to solipsisms; in the latter case, we can hope to recover historical
and quasiclassical inferences in an entirely straightforward way.
There is no methodological or scientific reason to prefer the former
position: it is tenable, but it certainly occupies no distinguished
high ground.
Although alternative dynamical theories are very interesting, it
remains a sensible null hypothesis to suppose that the dynamical
principles of quantum theory are correct.  
However, having made that
supposition, and under the assumption that the consistent histories
formalism defines the correct framework for making sense of quantum
theory, it is no less wild to hypothesise
that the formalism of should be augmented, at a fundamental level,
by a theory of experience than that it needs a set selection
criterion.  If quantum dynamics are fundamentally correct, and the
consistent histories formalism provides the correct physical interpretation,
then it seems that one hypothesis or the other must be right: at any
rate, no others have yet been suggested.
Both hypotheses go beyond quantum theory as it is
currently understood; neither is strongly supported by current science.

One sometimes encounters the following objection.  Suppose that a good set
selection hypothesis were found.  Would we not still ultimately need a
theory of experience?  If so, what would have we gained?  The answer, of
course, is that a theory of experience would still be sought ---
consciousness is a deep and fascinating problem in its own right.
What would be gained, though, by the selection of a quasiclassical history
--- {\it the} quasiclassical history of the world --- and the
assumption of psychophysical parallelism, is the knowledge that any
theory of experience would necessarily be formulated in terms of
quasiclassical variables.  The task of forming such a theory
would be simplified enormously over that of forming a theory in the
unaugmented consistent histories formalism (which is not to say that it would
be easy even were the variables known, it just speaks to the
extraordinary difficulties of forming the
theory otherwise). Then, making the weak assumption that
whatever the theory of experience is, it must be the case that
it would describe our experience (largely) mirroring real physical events
--- i.e. we {\it see} a table (usually) because there {\it is} a table
there --- we would not need the details
of the theory of experience to be able to predict the appearance to us of a
quasiclassical world.
This last assumption amounts to no more than the standard and well-supported
hypothesis that the quasiclassical variables in our brains are
correlated, by mechanisms familiar to neuroscientists, with those in
the external world.

Another objection occasionally raised is that it remains conceivable
that the operators relevant to our experience are {\it not}
quasiclassical, perhaps because they describe truly microscopic events
at the subneuronal level.\refs{\penrosebook}
There is no neuroscientific evidence for this hypothesis, 
and even if it were correct, the dichotomy would remain: either a 
theory of experience or a theory of reality would be needed, 
though a purely quasiclassical set selection rule would of 
course be excluded,

We would like to conclude by drawing attention to
d'Espagnat's admirable critique,\refs{\despagnat} which covers the
earlier papers of Griffiths and Omn\`es.
D'Espagnat examines the question of whether the consistent
histories formalism allows a realistically interpretable
local formulation of quantum mechanics, and
finds that it cannot, despite the impression which
might be gained by superficial readings.  To quote his
conclusion:\medskip\noindent
\medskip\noindent
``$\ldots$ it must be granted that several of the interpretative
comments the quoted authors make of their theories stand quite at odds
with the main conclusions reached here.  Indeed, while these authors
do not actually {\it say} their theories are realistically
interpretable,
they somehow give at various places the impression that they {\it
mean} just precisely that.  Such a somewhat disquieting state of
affairs seems to indicate that we physicists still have efforts to
make before we succeed in imparting to the {\it words} we use (and
especially to the nonoperationally defined ones) a strictness of
meaning comparable with the strictness of our mathematical
manipulations.
This will presumably only be achieved when we have convinced ourselves
that it is impossible to freely switch between an ontological and a
purely operationalist usage of such words as `have,' `is,'
`objective,' and the rest.''\medskip\noindent
\medskip\indent
We find ourselves very much in sympathy with d'Espagnat.
\bigskip\bigskip\centerline{{\bf Acknowledgements}}\nobreak
We thank the Aspen Institute for their hospitality
in the formative stages of this work.
F.D. was supported by DOE and
NASA grant NAGW-2381 at Fermilab and by NSF grant no. PHY-9008502;
A.K. by a Royal Society University Research
Fellowship.
It is a pleasure to thank Murray Gell-Mann, Bob Griffiths and
Roland Omn\`es for invaluable discussions of their work.
We would particularly like to thank Jim Hartle for taking the
time to explain his ideas to us and for
his generous support and
constructive criticisms in many discussions over the last two years.
We are very grateful to Trevor Samols for many helpful discussions
and to Shelly Goldstein for many thoughtful criticisms and insights.
We have also been greatly helped by discussions with
Andy Albrecht, Arley Anderson, Jeremy Butterfield, Matthew Donald,
Jerome Gauntlett, Nico Giulini, Peter Goddard, 
Jonathan Halliwell, Chris Isham, Achim Kempf, Raymond LaFlamme,
Klaas Landsman, Noah Linden, Jim McElwaine, Don Page,
Michael Redhead, Lee Smolin, Rafael Sorkin, 
John Taylor, Bill Unruh, G\'erard Watts and Wojciech Zurek.
\listrefs

\end